\documentclass[aps,reprint,nofootinbib,twocolumn,superscriptaddress]{revtex4-1}
\usepackage{subfigure}
\usepackage{amsmath, amssymb,graphicx,color,bm,epstopdf}
\usepackage[dvipsnames]{xcolor}
\usepackage{bbold}
\usepackage{braket}
\usepackage{comment}
\usepackage{natbib}
\usepackage{slashed}

\newcommand{\be}{\begin{equation}}
\newcommand{\e}{\end{equation}}
\newcommand{\beml}{\begin{subequations}}
\newcommand{\eml}{\end{subequations}}
\newcommand{\beq}{\begin{eqnarray}}
\newcommand{\eq}{\end{eqnarray}}
\newcommand{\ba}{\begin{array}}
\newcommand{\ea}{\end{array}}
\newcommand{\bpm}{\begin{pmatrix}}
\newcommand{\epm}{\end{pmatrix}}
\newcommand{\bc}{\begin{cases}}
\newcommand{\ec}{\end{cases}}

\usepackage{placeins}
\usepackage{float}
\usepackage{upgreek}

\usepackage{todonotes}

\usepackage{epsfig}

\DeclareMathAlphabet{\oldcal}{OMS}{cmsy}{m}{n}
\newcommand{\bigo}[1]{\oldcal{O}\left(#1\right)}

\definecolor{amendments}{rgb}{0.0, 0.0, 0.7}

\usepackage[utf8]{inputenc}
\begin{document}
\title{Critical behavior of weakly disordered Ising model: Six-loop $\sqrt \varepsilon$ expansion study}

\author{M.\,V.\,Kompaniets}
\affiliation{Saint Petersburg State University, 7/9 Universitetskaya Embankment, St. Petersburg, 199034 Russia}
\author{A.\,Kudlis}
\affiliation{ITMO University, Kronverkskiy prospekt 49, Saint Petersburg 197101, Russia}
\author{A.\,I.\,Sokolov}
\affiliation{Saint Petersburg State University, 7/9 Universitetskaya Embankment, St. Petersburg, 199034 Russia}

\begin{abstract}
The critical behavior of three-dimensional weakly diluted quenched Ising model is examined on the base of six-loop renormalization group expansions obtained within the minimal subtraction scheme in $4-\epsilon$ space dimensions. For this purpose the $\phi^4$ field theory with cubic symmetry was analyzed in the replica limit $n\rightarrow 0$. Along with renormalization group expansions in terms of renormalized couplings the $\sqrt{\varepsilon}$ expansions of critical exponents are presented. Corresponding numerical estimates for the physical, three-dimensional system are obtained by means of different resummation procedures applied both to the $\sqrt{\varepsilon}$ series and to initial renormalization group expansions. The results given by the latter approach are in a good agreement with their counterparts obtained experimentally and within the Monte Carlo simulations, while resumming of $\sqrt{\varepsilon}$ series themselves turned out to be disappointing.
\end{abstract}

\maketitle

\section{Introduction}
The study of critical properties of weakly disordered quenched systems is a subject of a great interest over the decades. Along with other theoretical approaches the renormalization group (RG) methods had been applied for description of critical behavior of such systems. The foundation of the field was laid in works of Harris and Lubensky \cite{PhysRevLett.33.1540,PhysRevB.11.3573}, Khmelnitskii \cite{Khmelnitskii1975}, and Grinstein and Luther \cite{PhysRevB.13.1329}. Within the RG approach some effective Hamiltonian based upon translationally invariant $\phi^4$ field theory was considered which is thermodynamically equivalent to the randomly diluted $m$-vector model. Of particular interest is randomly diluted Ising model (RIM), $m=1$, whose critical exponents, according to Harris criterion\footnote{This criterion states that if a pure system possesses positive heat capacity critical exponent $\alpha$ then the presence of impurities has to change its critical exponents.} \cite{Harris_1974}, should differ from those of pure Ising model and form special class of universality -- RIM class. In general, RG ideas and approaches look rather promising in studying wide variety of disordered and stochastic open systems~\cite{BM2015,BGM15}.     

As is well known, because of the degeneracy of equations on zeros of $\beta$-functions in the lowest-order, one-loop approximation non-trivial fixed point describing the random critical behavior is absent in this case. 
Solution of this problem was found by Khmelnitskii \cite{Khmelnitskii1975} and reexamined by Grinstein and Luther \cite{PhysRevB.13.1329} and Lubensky.
They showed that the series for coordinates of the random fixed point and critical exponents have to be in powers of $\sqrt{\varepsilon}$ instead of $\varepsilon$. Their results were obtained within the lower order approximation and no numerical estimates for critical exponents were found. Subsequently, two-loop and three-loop contributions were calculated \cite{AA76,PhysRevB.16.3987,SHALAEV1977} and twenty years later five-loop expansions were obtained  \cite{SHALAEV1997105,FOLKHOLYAR1999}. Unfortunately, due to dramatically irregular structure of the obtained expansions even in the highest-order -- five-loop -- $\sqrt{\epsilon}$ approximation no stable numerical estimates of critical exponents \cite{Janssen_1995} were found, in contradiction with the results obtained within other theoretical approaches and in real experiments. At the same time, in some papers (see, e.g. \cite{PhysRevB.61.15114,BLAVATSKA2003221}) the attempts to resum the initial RG expansions obtained via minimal subtraction (MS) scheme in terms of renormalized couplings without addressing $\sqrt{\epsilon}$ expansions were made. The issue of legitimacy of this approach was considered in \cite{Schloms_1987,SCHLOMS1989639}. The results obtained in \cite{Janssen_1995,PhysRevB.61.15114,BLAVATSKA2003221} turned out to be radically different from those given by $\sqrt{\epsilon}$ expansions themselves. These results within the corresponding error bars are in a fair agreement with the numbers obtained by means of other theoretical and numerical approaches.

Along with RG in $4-\varepsilon$ dimensions to describe the critical properties of systems belonging to RIM universality class the RG approach in three dimensions was applied. During almost twenty years two-loop \cite{JG83,Holovatch1992}, three-loop \cite{SS81,maiersokolov3loop3d,SHPOT1989474}, four-loop \cite{MSS89,Mayer_1989}, five-loop \cite{PS00} and six-loop \cite{PV00} RG expansions were calculated. The efforts undertaken within the higher orders proved fruitful. Despite the fact that the RG perturbative series seem to be not Borel-summable \cite{PhysRevB.36.2212,PhysRevB.49.12003,AMR00} numerical estimates for critical exponents characterizing RIM class of universality found by means of appropriate resummation procedures turned
out to be in a good agreement with the results obtained in physical experiments and by Monte Carlo simulations. Regarding the latter, there is an extensive discussion about identity of critical behavior of different randomly dilute models \cite{BFM98,CMPV03,Berche2004,HTPV07,PhysRevB.76.094402,PhysRevE.82.062101,
Theodorakis2011,PhysRevE.87.012132}. By means of the finite-size scaling technique the randomly site-diluted Ising model (RSIM) was considered in \cite{BFM98}. They showed numerically that critical exponents of such systems are dilution independent and differ from that of pure material. The authors and other researchers \cite{CMPV03,HTPV07} considering this model concluded basing upon the numerical evidences that RSIM has to enter RIM universality class. In other works the randomly bond-diluted Ising model (RBIM) was examined \cite{Berche2004,HTPV07}. As previously, the authors in both works came to the conclusion that the critical properties of RBIM has to be described also by RIM universality class. In addition, $\pm J$ Ising model \cite{PhysRevB.76.094402,PhysRevE.87.012132} and random-bond
one (RBIM) \cite{PhysRevE.82.062101,Theodorakis2011} demonstrated the same behavior at criticality. The numerical estimates of critical exponents from the above mentioned works including some experimental results \cite{B00} are collected in Table~\ref{tab:compare_crit_exp}. Let us say a few words about experimental activity in this area. To study the critical behavior of random Ising model in experiment some crystalline mixture of two compounds is used. One of them is anisotropic uniaxial antiferromagnet, for example, $FeF_2$ or $MnF_2$, while $ZnF_2$ plays a role of impurity. It should be noted that in experiments, as a rule, only effective critical exponents are measured because of multiple difficulties preventing approach of genuine asymptotic region. Collection of experimentally determined values of critical exponents for various dilute magnets may be found elsewhere \cite{FHY2003}.    
\begin{table}[t]
 \centering
    \caption{Numerical estimates of critical exponents $\nu$, $\eta$, and $\alpha$ for RIM universality class obtained by means of different theoretical approaches for different models of disorder and in experiments. The values of critical exponents for pure Ising model (the lowest line) were calculated within the conformal bootstrap approach.} %
    \label{tab:compare_crit_exp}
     \setlength{\tabcolsep}{3.4pt}
    \begin{tabular}{lllll}
      \hline
      \hline
      Method & Paper& $\nu$ & $\gamma$& $\alpha$ \\
      \hline
        3D RG: 2l & \cite{JG83}             &$0.67813$      &$1.336$      &$-0.03440$     \\
        3D RG: 3l & \cite{SHPOT1989474}     &$0.671$        &$1.328$      &$-0.013$     \\
        3D RG: 4l & \cite{MSS89}            &$0.670$        &$1.317$      &$-0.011$       \\
        3D RG: 4l & \cite{Mayer_1989}       &$0.6714$       &$1.321$      &$-0.013$       \\
        MS: 3l  &\cite{Janssen_1995}        &$0.666$        &$1.313$      &$0.002$             \\
        MC& \cite{BFM98}              &$0.6837(53)$   &$1.342(11)$  &$-0.051(16)$   \\
        3D RG: 5l & \cite{PS00}             &$0.671(5)$     &$1.325(12)$  &$-0.0125(80)$  \\
        MS: 4l  &\cite{PhysRevB.61.15114}   &$0.675$        &$1.317$      &$-0.026$             \\
        3D RG: 6l & \cite{PV00}             &$0.678(10)$    &$1.336(20)$  &$-0.034(30)$   \\
        Exp.          & \cite{B00}          &$0.69(1)$      &$1.370(29)$  &$-0.10(2)$     \\
        MC            & \cite{CMPV03} &$0.683(3)$     &$1.3421(61)$ &$-0.049(9)$    \\
        MS: 5l  &\cite{BLAVATSKA2003221}    &$0.708$        &$1.364$      &$-0.124$     \\
        MC        & \cite{Berche2004} &$0.68(2)$      &$1.336(39)$  &$-0.029(42)$  \\
        MC      &  \cite{PPVK2007}     & $0.693(5)$     & $1.342(7)$   & $-0.079(15)$  \\
        MC      & \cite{HTPV07}       &$0.683(2)$     &$1.3414(40)$ &$-0.049(6)$    \\
        ERG      & \cite{Tissier_2002}       &$0.67$     &$1.306$ &$-0.01$    \\
        MC&\cite{PhysRevB.76.094402} &$0.682(3)$ &$1.3394(60)$ &$-0.0460(90)$\\
        MC &\cite{PhysRevE.82.062101} &$0.6843(67)$   &$1.346(13)$  &$-0.0547(69)$  \\
        MC & \cite{Theodorakis2011}              &$0.685(7)$     &$1.345(15)$  &$-0.055(21)$  \\
        MC&\cite{PhysRevE.87.012132}&$0.6835(25)$&$1.345(11)$ &$-0.0505(75)$\\
        MS: 6l &TW&$0.675(19)$& $1.334(38)$ & $-0.025(57)$ \\
        \hline
        Pure IM&\cite{Simmons-Duffin2017}&$0.629971(4)$  &$1.237075(8)$& $0.110087(12)$\\
    \hline
    \hline
    \end{tabular}
\end{table}

The realization of the current work was motivated mainly by two reasons. The presence of six-loop RG expansions in three dimensions, along with new results obtained very recently for related models within various advanced approaches \cite{Antipin_2019,Stergiou_2019,Kousvos_2019,Nandi_2020,Shapoval_2020,Vigneshwar_2019}, and lack of six-loop series at $D=4-\varepsilon$ is the first factor. Serving as complementary tools different RG approaches usually allow to get more complete and reliable information about critical behavior of a system. Another reason is recently calculated RG functions for $n$-vector model with cubic symmetry in the six-loop approximation in the frame of MS scheme in $4-\varepsilon$ dimensions \cite{ADZHEMYAN2019332}. To perform these calculations the authors used results for $O(n)$-symmetric $\phi^4$ field theory obtained earlier \cite{KP17}. The critical behavior of the random Ising model is known to be described by the cubic model in the replica limit $n\rightarrow 0$. Thus with RG functions calculated in \cite{ADZHEMYAN2019332} in hand we can extend the five-loop results \cite{SHALAEV1997105,FOLKHOLYAR1999} to the next perturbative order.

The paper is organized as follows. In Section 2 we present the model and the RG procedure employed. The quantities which we have to calculate are also defined here. In Section 3 all numerical results are presented. The coordinates of the  random fixed point are calculated by means of resummation of corresponding six-loop $\sqrt{\epsilon}$ expansions and via finding of numerical solutions of equations on zeros of $\beta$ functions directly at $\epsilon$=1 without addressing $\epsilon$ expansion. The estimates of critical and correction-to-scaling exponents for RIM class of universality are also presented here.
At the end we draw some conclusion.

\section{The model and the renormalization procedure}

As was discussed in the previous section the replica method enables one to study the critical behavior of RIM basing upon effective field theory with cubic symmetry~\cite{PhysRevB.13.1329}. It is worth mentioning that RG methods do not fix the concentration of impurities where the critical behavior corresponding to RIM starts to realize. Therefore further we assume the impurity concentration to be very (infinitely) small. Relevant Landau-Wilson action is as follows:
\begin{eqnarray}
&&S_d = \int d\vec{x}\Biggl\{\frac{1}{2} \Big[\left[\partial \varphi_{0}\right]^2+ \psi(x)\varphi_{0}^2\nonumber\\ &&\qquad\qquad\qquad\qquad\qquad+m_0^2\varphi_{0}^2\Big]+ \frac{1}{4!} g_{0}\left[\varphi_{0}^2\right]^2\Biggr\},\label{H_d}
\end{eqnarray}
where $\psi(x)$ -- random field with Gaussian distribution, $m_0$ -- a bare mass being proportional to $(T-T_0)$, while $T_0$ -- a mean-field critical temperature. The replica technique allows to reformulate the problem for impure system to that for the model with cubic symmetry described by the action 
\begin{eqnarray}
&&S = \int d\vec{x}\Biggl\{\frac{1}{2} \left[\left[\partial \varphi_{0\alpha }\right]^2+ m_0^2  \varphi_{0\alpha }^2\right] \nonumber\\ &&\qquad\qquad\qquad\qquad+ \frac{1}{4!} g_{01}\varphi_{0\alpha}^2\varphi_{0\beta}^2+\frac{1}{4!}g_{02}\varphi_{0\alpha}^4 \Biggr\}, \label{H}
\end{eqnarray}
where $\varphi_{0\alpha}$ is $n$-component bare field ($\alpha$'s replica of a scalar field), $g_{01}$ and $g_{02}$ are bare couplings. The action describes RIM in the limit $n\rightarrow 0$ and under conditions $g_{10}<0$, $g_{02}>0$. 

In this work we address the field-theoretical RG approach in spatial dimensionality $D = 4-\varepsilon$. The model~\eqref{H} is known to be multiplicative renormalizable. All its RG functions for arbitrary $n$  were recently calculated by the authors via MS renormalization scheme in the six-loop approximation~\cite{ADZHEMYAN2019332}. The details related to the renormalization procedure can be found there. Supposing that the replica symmetry is preserved, in order to approach corresponding limit -- $n\rightarrow 0$ -- we just equate, as usually, $n$ to zero in all series for $\beta$-functions and anomalous dimensions $\gamma_{m^2}$ and $\gamma_{\varphi}$. Before calculating of critical exponents characterizing RIM universality class one needs to find coordinates of the random fixed point. It can be done by solving the following system of equations:
\begin{eqnarray}\label{beta_eq_zero}
\beta_i(g_1^*,g_2^*,\varepsilon) = 0  \quad i= 1,2.
\end{eqnarray}
The stability of found fixed point $(g_1^*,g_2^*)$ is determined by the following matrix:
\begin{eqnarray}\label{stab_matrix}
\Omega=\begin{pmatrix}
\dfrac{\partial\beta_1(g_1, g_2)}{\partial{g_1} } & \dfrac{\partial\beta_1(g_1, g_2)}{\partial{g_2}}\\[1.4em] \dfrac{\partial\beta_2(g_1, g_2)}{\partial{g_1}} & \dfrac{\partial\beta_2(g_1, g_2)}{\partial{g_2}}
\end{pmatrix}.
\end{eqnarray}
The fixed point is stable if the eigenvalues $\omega_1$, $\omega_2$ of matrix \eqref{stab_matrix} are positive. Approaching the critical temperature the renormalized couplings reach their fixed point values which determine the critical exponent values via the following relations:
\begin{eqnarray}
&\alpha = 2-\dfrac{D}{2+\gamma_{m^2}^*} , \ \ \beta  = \dfrac{D/2 - 1 + \gamma_{\varphi}^*}{2 + \gamma_{m^2}^*}, \ \ \gamma  = \dfrac{2 - 2\gamma_\varphi^*}{2 + \gamma_{m^2}^*},& \nonumber \\
&\eta = 2\gamma_\varphi^*, \ \ \nu = \dfrac{1}{2+\gamma_{m^2}^*}, \ \ \delta  = \dfrac{D+2-2\gamma_{\varphi}^*}{D-2+2\gamma_{\varphi}^*}.& \label{abgd}
\end{eqnarray}
where $\gamma_{m^2}^*\equiv\gamma_{m^2}(g_1^*,g_2^*)$ and $\gamma_{\varphi}^*\equiv\gamma_{\varphi}(g_1^*,g_2^*)$.
The critical exponents are related to each other by well-known scaling relations and only two of them may be referred to as independent. 

\section{Numerical results}
As was already mentioned, in order to extract the numerical estimates from RG expansions obtained within MS scheme we can proceed in two different ways. First, which is traditional for this approach, consists of solving the equations \eqref{beta_eq_zero} by means of iterative calculations of $\sqrt{\epsilon}$ expansions for fixed point coordinates and subsequent substitution of this expansion to the series for critical exponents as functions of renormalized couplings. Although a priori, even with use of various resummation procedures, we do not expect getting reliable numbers following this approach, we present here, for completeness, all the $\sqrt{\epsilon}$ expansions and corresponding numerical estimates for the quantities of interest. Alternatively, and here we pin our hopes, one can analyze initial RG expansions obtained within MS scheme in three dimensions, without addressing the $\sqrt{\epsilon}$ expansions. From the very beginning in the series for $\beta$-functions we put $\epsilon=1$, that corresponds to $D=3$. After this we find coordinates of the random Ising fixed point solving the equations~\eqref{beta_eq_zero} numerically. The obtained numbers will be then substituted into the series for physically interesting quantities. It is implied that all the expansions including those for $\beta$-functions have to be treated by means of various resummation techniques.

\subsection{RG expansions}
The RG functions of the model~\eqref{H} in the replica limit $n\rightarrow 0$ obtained within MS scheme in the six-loop approximation with coefficients in analytical form can be found in Supplementary Materials as \textit{Mathematica}-file (\textit{rg\_expansions\_im\_is.m}).
The calculated coefficients are in a full agreement with the five-loop results~\cite{SHALAEV1997105}. With RG expansions in hand we can calculate all the quantities of interest. In the next section we will solve equations \eqref{beta_eq_zero} within two ways just described.

\subsection{Fixed-point coordinates}
\subsubsection{$\sqrt{\varepsilon}$ analysis}
As was already said, because of the degeneracy of equations \eqref{beta_eq_zero} in the lowest-order, one-loop RG approximation we are forced to work with $\sqrt{\varepsilon}$ series instead of traditional $\varepsilon$ expansions. For the fixed point coordinate $g_1$ the series with coefficients in analytical form is presented in Supplementary Materials as \textit{Mathematica}-file (\textit{fp\_coordinates\_im\_is.m}). To get an idea about the expansion coefficients behavior we present here the same series with coefficients in decimals:
\begin{eqnarray}\label{g1_num_sqrt_eps_exp}
&&g_1^*=-0.504695\varepsilon^{1/2}+0.595075\varepsilon-0.794926\varepsilon^{3/2}\nonumber\\
&&\qquad\qquad-0.581659\varepsilon^{2}-1.02342\varepsilon^{5/2}+\bigo{\varepsilon^{3}}. \ \ \ 
\end{eqnarray}
\noindent Since the expansion has obviously irregular structure it would be naive to wait for proper numerical estimate for this coupling even after applying advanced resummation procedures. To support our concerns, keeping in mind the lack of knowledge of asymptotic behavior for $\sqrt{\varepsilon}$ expansions, we construct Pad\'e--Borel--Leroy (PBL) triangle (Table \ref{pblg1}) for \eqref{g1_num_sqrt_eps_exp} where for all $\sqrt{\varepsilon}$ series before resummation we do the variable change $\varepsilon \rightarrow \varepsilon t^2$ and put $t=1$ at the end of calculations. The description of used algorithm -- determination of fitting parameter $b$ -- can be found in \cite{ADZHEMYAN2019332}.  
\begin{table}[h!]
\centering
\caption{Pad\'e-Borel-Leroy estimates of $g_1^*$ obtained from six-loop $\sqrt{\varepsilon}$ expansion via three the most stable approximants -- [4/1], [2/3], and [2/2] -- under the optimal value of the shift parameter $b_{opt} = 9.82$. Empty boxes correspond to the approximants spoiled by dangerous poles. Simple Pad\'e approximants are much worse, Borel-Leroy transformation and the tuning of fitting parameter $b$ notably improved the convergence of estimates.} 
\label{pblg1}
\renewcommand{\tabcolsep}{0.175cm}
\begin{tabular}{{c}|*{5}{c}}
\hline
\hline
$M \setminus L$ & 1 & 2 & 3 & 4&5 \\
\hline
0 & $-$0.5047 & 0.09038 & $-$0.7045 & $-$1.286 & $-$2.310  \\
1 & $-$0.2372 & $-$0.2435 & - & $-$0.2290 & \text{} \\
2 & $-$0.2439 & $-$0.2371  & - & \text{} & \text{} \\
3 & - & $-$0.2408 & \text{} & \text{} & \text{} \\
4 & - & \text{} & \text{} & \text{} & \text{} \\
\hline
\hline
\end{tabular}
\end{table} 

After averaging over the most reliable approximants we obtain: 
\begin{equation}
    g^*_{1,\sqrt{\varepsilon}}=-0.236(15).
\end{equation}

The $\sqrt{\varepsilon}$ expansion for the second coordinate $g_2$ with analytically expressed coefficients is presented in the same file. Its counterpart in decimals is
\begin{eqnarray}
&&g_2^*=0.672927\varepsilon^{1/2}-0.567019\varepsilon+0.92997\varepsilon^{3/2}\nonumber\\
&&\qquad\qquad+1.27497\varepsilon^{2}+1.03926\varepsilon^{5/2}+\bigo{\varepsilon^{3}}. \ \ \ 
\end{eqnarray}
The situation here is much worse and there is no possibility to construct more or less suitable PBL triangle for any values of fitting parameter $b$. That is why in the next section we address the alternative way of processing of RG series calculated within MS scheme.

\subsubsection{Summation of RG series for $\beta$-functions under $\varepsilon$=1}

Let us describe in details the strategy by which the series for $\beta$-functions are processed. We put $\varepsilon=1$ from the very beginning what corresponds to $D=3$. If the information about the asymptotic behavior of the series coefficients is known the range of resummation procedures can be supplemented with the most effective ones. For $\varphi^4$ field theory with cubic symmetry the large-order behavior of expansions coefficients is known. One can expect that by analytic continuation the results can be extended to $n=0$. However, it is not the case here. As was shown in~\cite{PhysRevB.36.2212,PhysRevB.49.12003} and extensively discussed in~\cite{PV00}, RG series for zero-dimensional RIM are not Borel summable and there are no any evidences to believe that the lack of Borel summability does not reveal itself in higher spatial dimensions.

In order to overcome this problem and resum the divergent expansions, we address the recipe applied in the course of 3D RG analysis~\cite{PV00}. The corresponding strategy is as follows. First, one has to reexpand the RG expansions for $\beta$-functions in terms of isotropic coupling $g_1$ with coefficients depending on cubic coupling $g_2$:
\begin{align}
    \beta_i(g_1,g_2)&=\sum\limits_{k=1}^{\infty}\beta_{k}(g_2) g_1^k,\label{beta_reexpanded_1_2}\\
    \beta_{k}(g_2)&=\sum\limits_{l=1}^{\infty}\beta_{k,l}g_2^l.\label{reexpanded_coef}
\end{align}
According to the results obtained in~\cite{AMR00}, the expansion ~\eqref{reexpanded_coef} and corresponding series~\eqref{beta_reexpanded_1_2} at fixed $g_2$ are already Borel summable. Thus, as the first step we have to resum all the coefficients $\beta_{k}(g_2)$ as the series in $g_2$, after that one can proceed to the resummation of expansions in terms of $g_1$. Due to the existence of Borel summability for $g_2$ series, we can apply both PBL and conform-Borel (CB) resummation techniques. We present here the numerical results obtained within both approaches. Since the detailed description of mentioned resummation procedures was given in many papers (see, e. g.,~\cite{PhysRevB.21.3976,GZ97,GZJ98,ZJ01,ZJ,PV02}) here we restrict ourselves with the working formulas and description of the algorithm of choosing the resummation parameters. 
First, we separate in series for $\beta$-functions their "loop" (perturbative) parts. Within the normalization we adopt, in the critical point each part has to coincide with value of the corresponding random Ising fixed point coordinate:
\begin{eqnarray}\label{app_formula_cm_last1}
    \beta_i(g_1,g_2)&=&\sum\limits_{k,l=1}^{\infty}\beta_{k,l} g_1^k g_2^l,\\ \beta_{i}^{\,loop}(g_1,g_2)&=&\beta_i(g_1,g_2)+g_i, \ \ \ i=1,2. 
\end{eqnarray}
Next, we reexpand the coefficients of the loop parts in accordance with expressions~\eqref{beta_reexpanded_1_2} and~\eqref{reexpanded_coef}. After that we resum the coefficients $\beta_{k}(g_2)$ by means of the one of two techniques mentioned above. Having obtained the values of coefficients $\beta_{k}(g_2)$ the PBL resummation procedure is applied to series in $g_1$. The main challenge here is to determine the optimal values of resummation parameters. These parameters have no any physical meaning, therefore it is quite natural to take such their values under which true physical quantities (critical exponents, etc.) would be less sensitive to their variation. The corresponding reasonable ranges where we analyze the resummation parameters are as follows. The \textit{Leroy} parameters $b_1$ and $b_2$ are being varied through $b^{PBL}\times b^{PBL}$, where
$b^{PBL}_{range}=[0.0,15.0,0.05]$ (90601 points) for PBLPBL resummation procedure. In case of CBPBL technique $b_1$ varies within the same range with different step $b^{PBL_{CB}}_{range}=[0.0,15.0,0.25]$, while $b_2$ is being varied via $b^{CB}_{range}=[0.0,5.0,0.25]$. The range for $\lambda$ parameter which is relevant only for CBPBL resummation procedure is $\lambda_{range}=[0.0,2,0.1]$ (26901 points). 

Thus, in the case of using the PBLPBL technique for $\beta_{k}(g_2)$ series we take the first point from $b^{PBL}_{range}\times b^{PBL}_{range}$ and calculate the resummed $\beta$-functions by means of the following expression:
\begin{eqnarray}\label{app_formula_cm_last2}
&&\beta^{\,R,loop}_{i}(g_1,g_2)\approx\beta_{i,L,M,S,J}^{\,loop,N}(g_1,g_2,b_1,b_2)\nonumber\\
&&\quad=\int\limits_{0}^{\infty}dt_1^{} \, t_1^{b_1} e^{-t_1}P_{L,M}\left[\sum\limits_{k=0}^{N}\frac{\beta_k^{S,J}(g_2,b_2)}{\Gamma(k+1+b_1)}\left(g_1t_1\right)^k \right],\quad \ \ 
\end{eqnarray}
where the coefficients $\beta_k^{S,J}(g_2,b_2)$ are computed as:
\begin{equation}
\beta_k^{S,J}(g_2,b_2)=\int\limits_0^{\infty}\,dt^{}_2 t_2^{b_2} e^{-t_2}P_{S,J}\left[\sum\limits_{l=0}^{N-k}\frac{\beta_{k,l} \left(g_2t_2\right)^l}{\Gamma(l+1+b_2)}\right].
\end{equation}
The operators $P_{L,M}$ and $P_{S,J}$ generate the Pad\'e approximants for the Borel--Leroy image of the corresponding series. The choice of values $L$, $M$, $S$, and $J$ will be discussed later.

In the case one addresses an alternative, CB resummation technique in order to tackle the $\beta_{k}(g_2)$, the resummation parameter space becomes three-dimensional. Therefore, we take a point from $b^{PBL_{CB}}_{range}\times b^{CB}_{range}\times \lambda_{range}$ and work with the following system of equations:
\begingroup
\allowdisplaybreaks
\begin{widetext}
\begin{eqnarray}\label{app_formula_cm_last2123}
&\beta^{\,R,loop}_{i}(g_1,g_2)\approx\beta_{i,L,M}^{\,loop,N}(g_1,g_2,b_1,b_2,\lambda)=\int\limits_{0}^{\infty}dt^{}_1 \, t_1^{b_1} e^{-t_1}P_{L,M}\left[\sum\limits_{k=0}^{N}\dfrac{\beta_k(g_2,b_2,\lambda)}{\Gamma(k+1+b_1)}\left(g_1t_1\right)^k \right],&\\
&\beta_k(g_2,b_2,\lambda)=\int\limits_{0}^{\infty}dt_2 \ t_2^{b_2} e^{-t_2}\left(\dfrac{g_2t_2}{w(g_2t_2)}\right)^{\lambda}\sum\limits_{l=0}^{N-k}W_{l,b_2,\lambda}[\beta_k(g_2)](w(g_2t_2))^l,\label{beta_k_cb}&\\ 
&\left(\dfrac{g_2}{w(g_2)}\right)^{\lambda}\sum\limits_{l=0}^{N-k}W_{l,b_2,\lambda}[\beta_k(g_2)](w(g_2))^l=\sum\limits_{l=0}^{N-k}\dfrac{\beta_{k,l}}{\Gamma(l+b_2+1)}g_2^l+\bigo{g_2^{N-k+1}},&\\
&w(g_2)=\left(\sqrt{1+g_2}-1\right)\left(\sqrt{1+g_2}+1\right)^{-1},&
\end{eqnarray}
\end{widetext}
\endgroup
\noindent where the form of function $w(g_2)$ is known from the large-order behavior analysis for the pure Ising model.

Having obtained the resummed $\beta$-functions which depend on the coordinates $g_1$ and $g_2$ one can find a root (roots) of the following system of equations:
\begin{eqnarray}\label{working_formula_for_zeros_betas}
    F_i[g_1^*,g_2^*]&=&0,\\
    F_i[g_1,g_2]&=&\beta_{i}^{\,R,loop}(g_1,g_2)-g_i, \ \ i=1,2.
\end{eqnarray}
within some apriori known range of expected values of coordinates determined, say, from the PBL analysis of $\sqrt{\varepsilon}$ series. If there is no root we discard the currently considered point $(b_1,b_2)$ (or $(b_1,b_2,\lambda)$) and move to the next one. Otherwise we begin to analyze the stability of the result obtained. In fact, the found candidate for the fixed point coordinate is a function of the resummation parameters and also depends on the choice of the Pad\'e approximants used. At this stage, we start to measure the spreading of fixed point coordinates estimates in the vicinity of the corresponding point in resummation parameters space. 

When PBL technique is used for both couplings, the procedure is as follows. For particular values of $b_2$ we construct a set of approximants $P_{S,J}$, which includes those of two highest orders available without poles on the positive real axis. Among these approximants we choose three most close to each other. On the base of this new set we find the values for $\beta_i$ coefficients. Having obtained the new series in terms of $g_1$ we construct the similar set of approximants for this new expansion and again determine a set consisted of three closest to each other approximants. Basing upon this sample we obtain the value of candidate to the fixed point and take the standard deviation as the measure of spreading $\Delta_{PBL}(b_1,b_2)$. We add the obtained information to the data and start to analyze the next point from $b^{PBL}_{range}\times b^{PBL}_{range}$. 

In case of addressing the CB technique for resummation of $\beta_i(g_2)$, we take a point $(b_2,\lambda)$ and calculate these coefficients according to equation~\eqref{beta_k_cb}. With the resummed coefficients we start to analyze new series in terms of $g_1$ by means of PBL technique. For this purpose we vary the value of parameter $b_1$ within the suggested range and for each point $b_1$ from $b^{PBL_{CB}}_{range}$ construct the set of the most stable Pad\'e approximants. We define the optimal value of $b_1$ as the point, where the standard deviation calculated on the basis of the mentioned sample would be the minimal. After that we start to analyze the stability of the obtained estimates for the fixed point coordinates. For this purpose we use the same approach as in PBL case. Having completed the first cycle, we add all information related to this point -- $\{(b_1,b_2,\lambda),\{g_1^*,g_2^*\},\Delta_{CB}(b_2,\lambda)\}$ -- to the database and then repeat all the above steps for the whole grid of resummation parameters.

In both cases the minimal value of the $\Delta$, in fact, can be considered as an indicator that analyzed functions -- $\beta^{\,R,loop}_{i}(g_1^*,g_2^*)$ --  achieved a plateau. After analyzing the whole grid, we find the point with the minimal value of $\Delta$. It is natural to take as a final estimate of the fixed point coordinate the corresponding value of $\{g_1^*(b_1^*,b_2^*),g_2^*(b_1^*,b_2^*)\}$ (or $\{g_1^*(b_1^*,b_2^*,\lambda^*),g_2^*(b_1^*,b_2^*,\lambda^*)\}$). However, in order to exclude the possibility of accidentally extreme deviation we perform the following steps.
\begin{figure}[h!]
    \centering
    \includegraphics[width=0.46\textwidth]{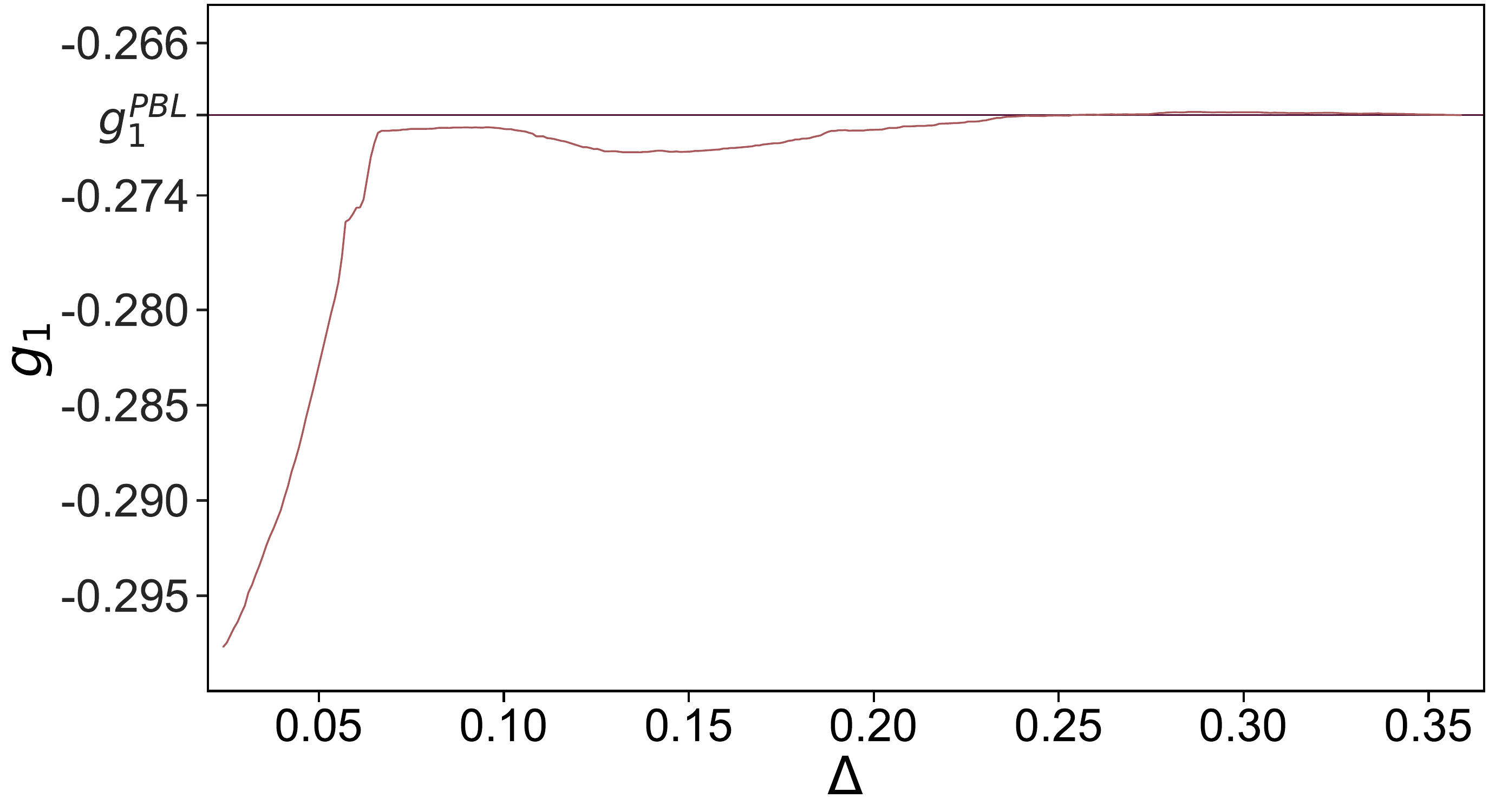}
    \caption{The dependence of $g_1^*$ PBLPBL estimate on maximally allowed spreading of points entering the sample on the basis of which the mentioned estimate was obtained. The plateau value is $-0.2698$.}
    \label{fp_PBLPBL_g1_estimate}
\end{figure}
\begin{figure}[h!]
    \centering
    \includegraphics[width=0.46\textwidth]{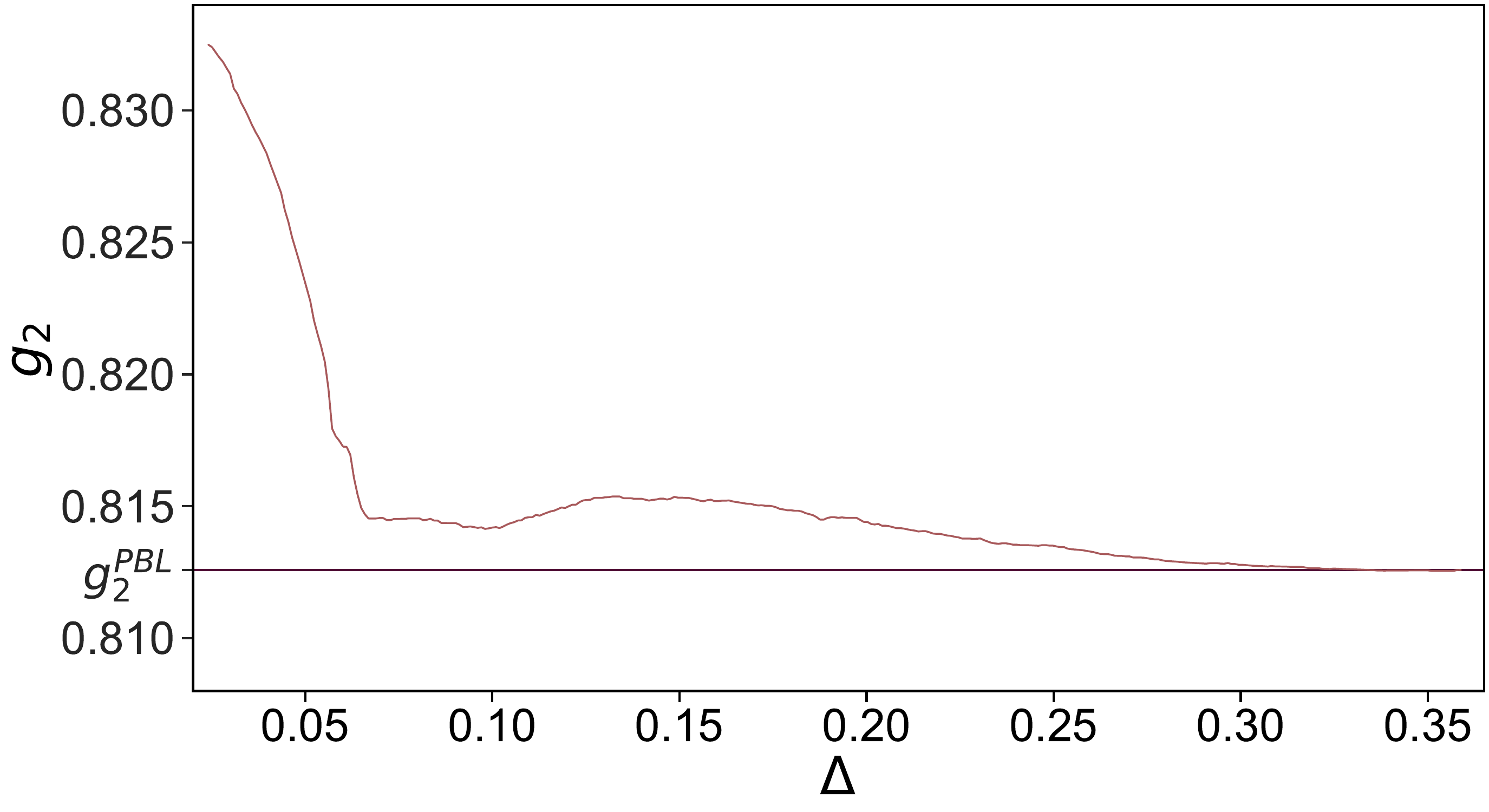}
    \caption{The dependence of $g_2^*$ PBLPBL estimate on maximally allowed spreading of points entering the sample on the basis of which the mentioned estimate was obtained. The plateau value is $0.8126$.}
    \label{fp_PBLPBL_g2_estimate}
\end{figure}
In both cases we analyze the evolution of working set with increasing of maximally allowed spreading for points entering it, including dependencies of the mean values and standard deviations.
The behavior of the mean value is shown in Fig.~\ref{fp_PBLPBL_g1_estimate} for $g_1$ coordinate and in Fig.~\ref{fp_PBLPBL_g2_estimate} for $g_2$ coupling when PBLPBL procedure is used.

Varying the maximal value of allowed spreading one can note that in PBLPBL case the estimates of discussed quantities achieve some plateau values. Here we stop at $\Delta=0.35$ which corresponds to sample consisted of $2777$ points from $4241$ found roots for $\beta$-functions on the basis of $90601$ points of parametric $b^{PBL}_{range}\times b^{PBL}_{range}$ space. 

If CBPBL technique is applied the total number of roots obtained on the basis of the mentioned parametric grid is $649$. Having performed the analogous steps as we do in PBLPBL case we do not obtain a similar fortunate behavior of estimates. We did not find any plateau. In such a situation we have to take into account all the points what results in larger uncertainty for CBPBL method.

Thus, our final estimates of the random fixed point coordinates obtained by means of PBLPBL and CBPBL resummation procedures are as follows:  
\begin{eqnarray}
&&g_{1,PBL}^*=-0.270(19), \quad g_{2,PBL}^*=0.813(26),\label{pblpbl_res_num_fp}\\
&&g_{1,CB}^*=-0.15(11), \quad \quad g_{2,CB}^*=0.66(9).\label{cbpbl_res_num_fp}
\end{eqnarray}
\begin{figure}[h!]
    \centering
    \includegraphics[width=0.46\textwidth]{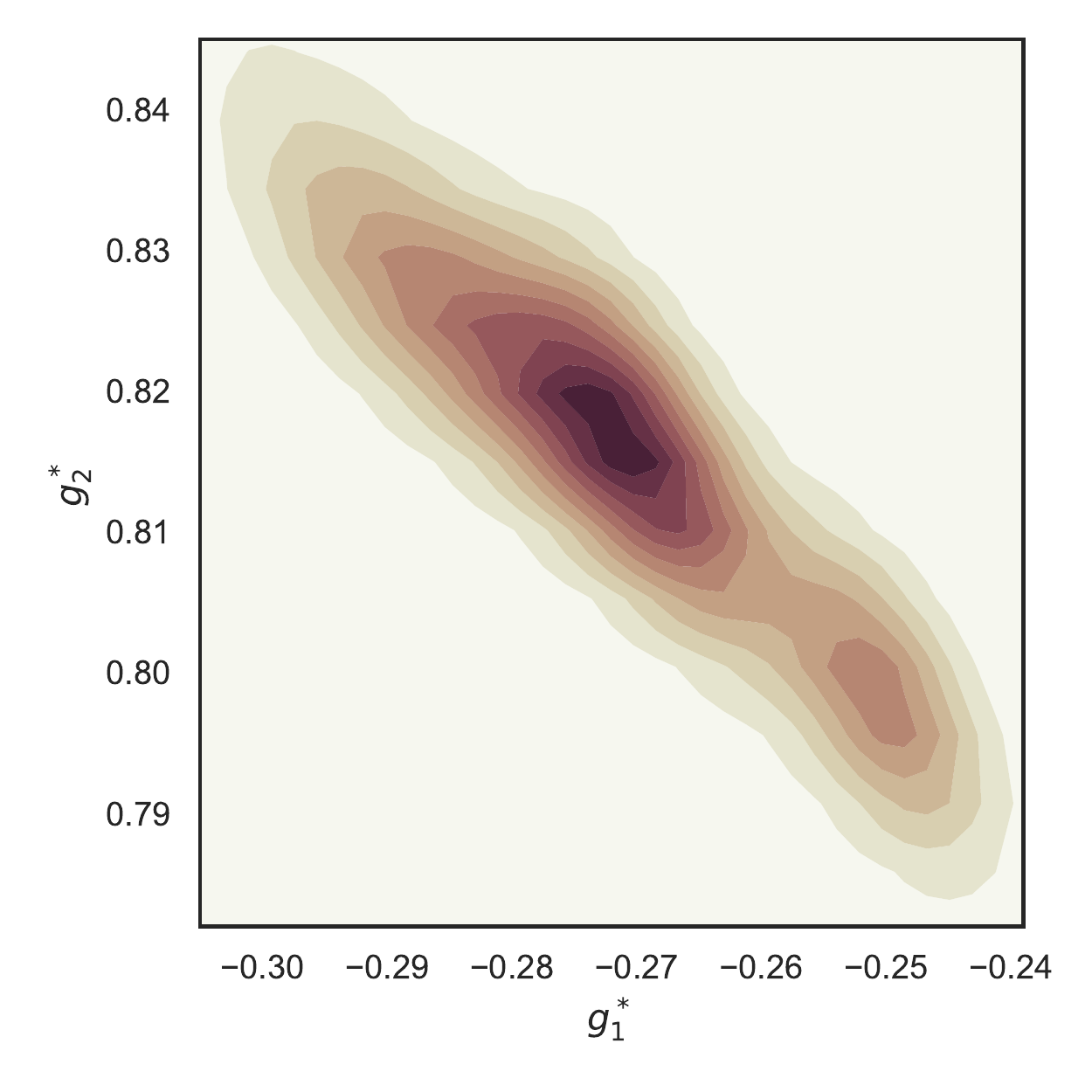}
    \caption{The distribution of estimates of fixed point coordinates when PBL technique is applied to resum coefficients $\beta_i(g_2)$.  The sample consists of 2777 unique points from 90601 possible candidates.}
    \label{fp_PBLPBL_fp}
\end{figure}
To make the results of our calculations more visual we plot  the distribution of random fixed point coordinates as functions of varying resummation parameters which is based on the sample chosen for extracting final numerical estimates in PBLPBL case. Corresponding histogram is presented in Fig.~\ref{fp_PBLPBL_fp}. 
It is worthy to note that the estimates of critical exponents obtained on the basis of PBLPBL results for random fixed point location will be considered as those of highest priority since PBLPBL resummation turned out to provide higher numerical accuracy.   

\subsection{Critical exponents}
We turn now to the obtaining of numerical estimates for the physically important quantities -- critical exponents. Here we limit ourselves by considering the most popular ones --  $\nu$ and $\gamma$.
Each exponent is calculated by means of resummation of corresponding $\sqrt{\varepsilon}$ expansion as well as by processing of initial RG series in the three-dimensional way. In case of $\sqrt{\varepsilon}$ series we use conventional PBL technique, while for resummation of expansions for critical exponents in terms of renormalized couplings we address the strategy similar to that used for finding of the fixed point coordinates. 

\subsubsection{Critical exponent $\nu$}
The six-loop $\sqrt{\varepsilon}$ expansion for correlation length critical exponent $\nu$ with coefficients in analytical form as well as the series for $\eta$ and $\gamma$ and correction-to-scaling exponents are presented in Supplementary Materials as \textit{Mathematica}-file (\textit{critical\_exponents\_im\_is.m}). Up to the five-loop contribution -- $\bigo{\varepsilon^{2}}$ -- all the coefficients coincide with those obtained earlier ~\cite{SHALAEV1997105}. In decimals this series reads
\begin{eqnarray}\label{sqrt_eps_num_nu_exp}
&&\nu=0.5+0.0841158\varepsilon^{1/2}-0.016632\varepsilon+0.0477535\varepsilon^{3/2}\nonumber\\
&&\qquad\qquad+0.272584\varepsilon^{2}+0.223298\varepsilon^{5/2}+\bigo{\varepsilon^{3}}.
\end{eqnarray}
Let us first construct PBL triangle for this expansion. In comparison with, for example, the critical exponent $\eta$ a situation here is a bit better and numerical estimates have been found to be stable enough.
\begin{table}[h!]
\centering
\caption{Pad\'e-Borel-Leroy estimate of critical exponents $\nu$ obtained from six-loop $\sqrt{\varepsilon}$ expansion under averaging through the highest-order and the most stable approximants under the optimal value of the shift parameter $b_{opt} = 10.53$.}
\label{pblnu}
\renewcommand{\tabcolsep}{0.165cm}
\begin{tabular}{{c}|*{6}{c}}
\hline
\hline
$M \setminus L$ & 0 & 1 & 2 & 3 & 4&5 \\
\hline
 0&   0.5 & 0.5841 & 0.5675 & 0.6152 & 0.8879 & 1.111 \\
 1&0.6034 & 0.5704 & 0.5796 & 0.5561 & 0.6045 & \text{} \\
 2&0.5583 & 0.5858 & 0.5649 & 0.5940 & \text{} & \text{} \\
 3&0.6765 & 0.5348 & 0.6329 & \text{} & \text{} & \text{} \\
 4&0.05334& 0.5543 & \text{} & \text{} & \text{} & \text{} \\
 5&1.266 & \text{} & \text{} & \text{} & \text{} & \text{} \\
\hline
\hline
\end{tabular}
\end{table}
The PBL triangle is presented in Table~\ref{pblnu}. After averaging over the most stable approximants the estimate for $\nu$ is as follows:
\begin{equation}
    \nu_{\sqrt{\varepsilon}}= 0.577(31).
\end{equation}
This number strongly differs from any known results obtained theoretically or within the experiments. Moreover, it is in  obvious contradiction with the theorem stating that for 3D RIM the specific heat exponent $\alpha = 2 - 3 \nu$ should be negative \cite{SS81,MaBook}.  

In a such situation we have to address an alternative approach of series treatment. With numerical estimates for the fixed point coordinates in hand we can extract the values of critical exponents by means of applying PBLPBL and CBPBL techniques to RG series in terms of couplings with important simplification of numerical analysis due to the fact that there is no need to find the FP coordinates. It is quite natural to use estimate  the~\eqref{pblpbl_res_num_fp} for FP coordinates when PBLPBL technique is used for resummation of critical exponent series, while the numbers ~\eqref{cbpbl_res_num_fp} should be employed when CBPBL procedure is applied. To get an idea about coefficients behavior of RG series we have to do the following change of variables: $g_i\rightarrow g_i^*z$, for $i=1,2$. Thus, for PBLPBL and CBPBL cases we have two alternative series for exponent $\nu$. They read:
\begin{eqnarray}\label{nu_rg_num_exp_pblpbl}
&&\nu_{PBL}=0.5+0.1583(72)z-0.0063(35)z^{2}+0.111(21)z^{3}\nonumber\\
&&\quad-0.351(83)z^{4}+1.51(45)z^{5}-7.5(2.7)z^{6}+\bigo{z^{7}},\quad\\
&&\nu_{CB}=0.5+0.145(25)z-0.012(12)z^{2}+0.100(68)z^{3}\nonumber\\
&&\quad-0.30(26)z^{4}+1.2(1.3)z^{5}-5.8(7.7)z^{6}+\bigo{z^{7}}.\quad
\end{eqnarray}
In contrast to~\eqref{sqrt_eps_num_nu_exp} these series possess a regular structure -- they are alternating and demonstrate apparently factorial growth of the coefficients. Let us resum them. In case of PBLPBL and CBPBL resummation techniques we address the formulas which are similar to~\eqref{app_formula_cm_last2} and ~\eqref{app_formula_cm_last2123} but with substitution, instead of couplings, their fixed point values $g_1^*$ and $g_2^*$ from~\eqref{pblpbl_res_num_fp} and~\eqref{cbpbl_res_num_fp} respectively. 
\begin{figure}[b!]
    \includegraphics[width=0.48\textwidth]{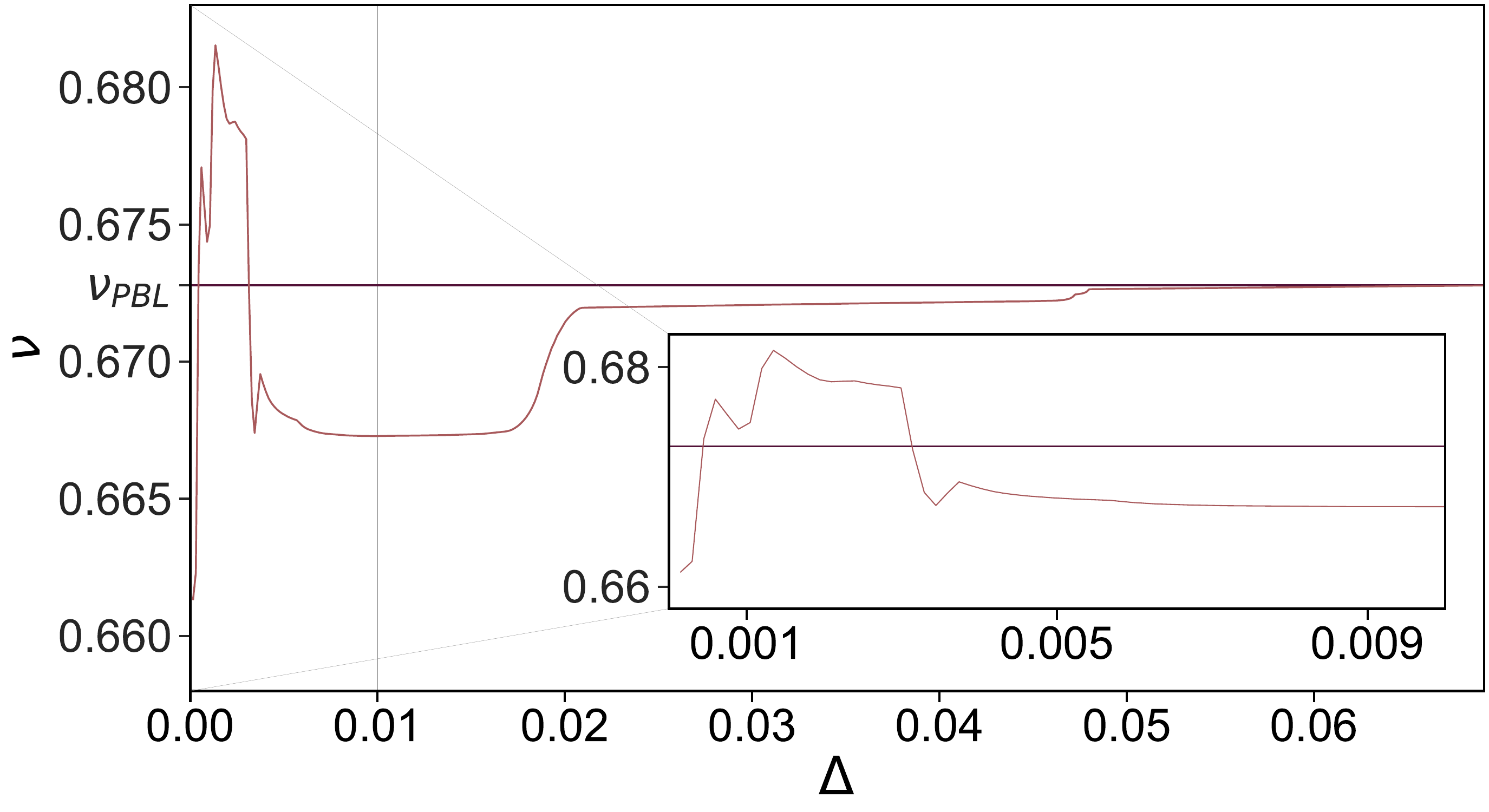}
    \caption{The dependence of critical exponent $\nu$ estimate on the maximally allowed value of spreading in the analyzed array. The plateau value is $0.6728$. 
    }
    \label{nu_PBLPBL_spre_fun}
\end{figure}
\begin{figure}[h!]
    \includegraphics[width=0.48\textwidth]{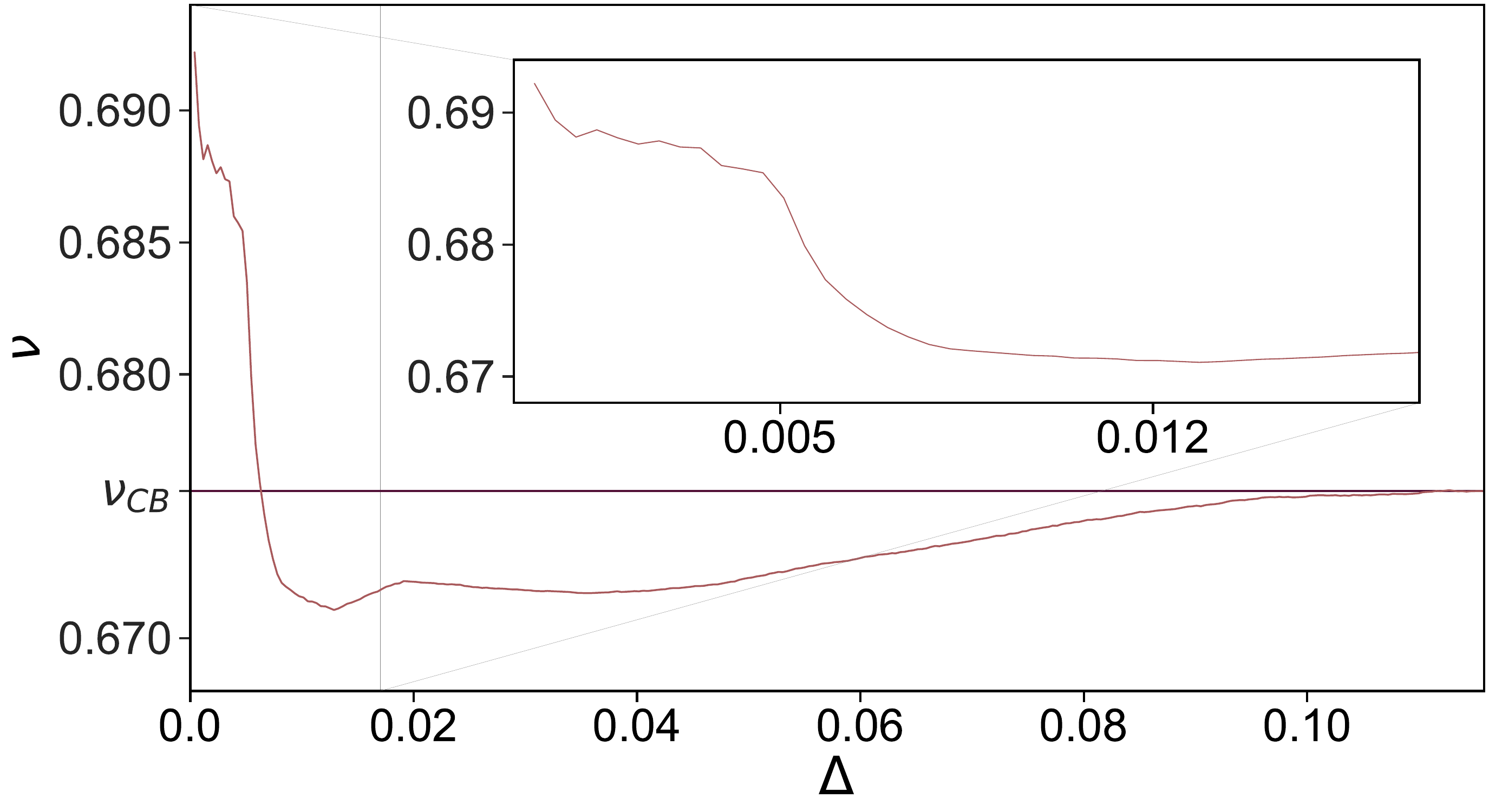}
    \caption{The dependence of critical exponent $\nu$ estimate on the maximally allowed value of spreading in the analyzed array. The plateau value is $0.6755$. 
    }
    \label{nu_CBPBL_spre_fun}
\end{figure}
As previously, we analyze how the number of points which enter the array, the mean value of $\nu$ and its standard deviation depend on the spreading. Based on the whole grid of obtained points 90601 (26901) and approximate absolute value of critical exponent $\nu$ ($\sim 0.65$)  we choose the sufficiently conservative strategy of spreading change, varying $\Delta$ from zero till the moderate measure when the mean value of the analyzed sample achieves some plateau. In case of PBLPBL and CBPBL resummation techniques, the results of analysis are as follows:
\begin{eqnarray}
     \nu_{PBL}=0.673(18), \qquad \nu_{CB}=0.676(19).
\end{eqnarray}
On the base of these numbers we accept the value
\begin{eqnarray}
     \nu=0.675(19).
\end{eqnarray}
as the final estimate for critical exponent $\nu$.
The corresponding dependencies of mean value on $\Delta$ for PBLPBL ans CBPBL cases are presented in Fig.~\ref{nu_PBLPBL_spre_fun} and~\ref{nu_CBPBL_spre_fun} respectively. 
\begin{figure}[h!]
    \centering
    \includegraphics[width=0.46\textwidth]{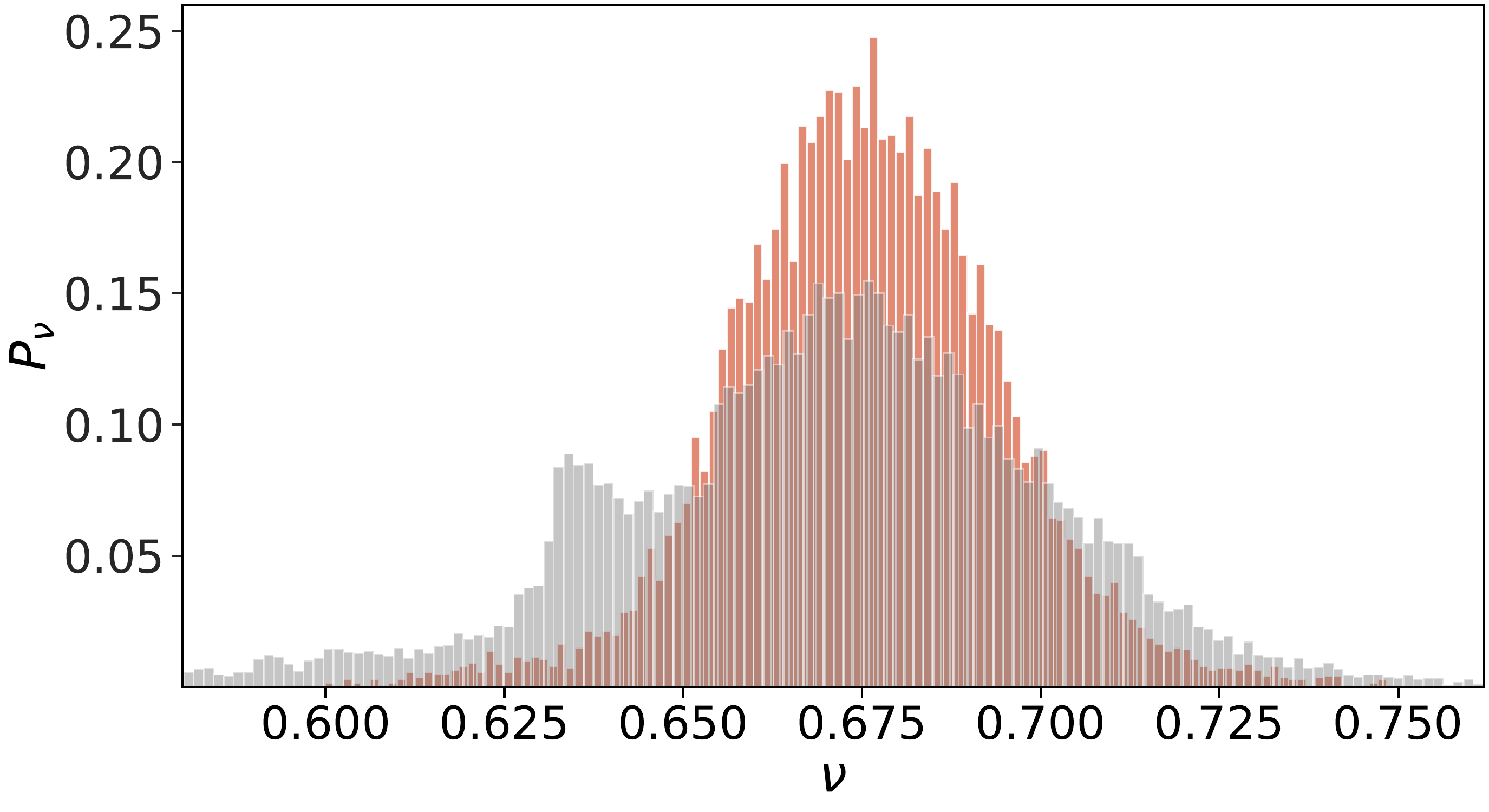}
    \caption{The distribution (red one or more "Gaussian-like") of estimates of critical exponent $\nu$ when CBPBL technique is applied in case of $\Delta=0.11583$.  
    The "gray" distribution corresponds to $\Delta=0.2$.}
    \label{nu_CBPBL_hist}
\end{figure}

One should note that in CBPBL case the estimates of standard deviation do not achieve the plateau within the moderate value of $\Delta$. However, we do not continue to increase the $\Delta$ because of the following reason. The further growth of $\Delta$ leads to appearing of "white noise" and extra peaks in addition to the main one in the distribution Fig.~\ref{nu_CBPBL_hist} what strongly spoils the accuracy.

\subsubsection{Critical exponent $\gamma$}
The $\sqrt{\varepsilon}$ expansion for susceptibility exponent $\gamma$ of RIM in decimals is
\begin{eqnarray}\label{gamma_sqrt_decimals}
&&\gamma=1 +0.168232\varepsilon^{1/2}-0.0285471\varepsilon+0.0788288\varepsilon^{3/2}\nonumber\\
&&\qquad\qquad+0.564505\varepsilon^{2}+0.440615\varepsilon^{5/2}+\bigo{\varepsilon^{3}}.\
\end{eqnarray}
The PBL triangle for this expansion is presented in Table \ref{pbl_gamma}. The numerical estimates for $\gamma$ resulting from this triangle is
\begin{equation}
    \gamma_{\sqrt{\varepsilon}}=1.172(55).
\end{equation}
As in the case of exponent $\nu$, the estimate for $\gamma$ obtained by means of resummation of $\sqrt{\epsilon}$ expansion is in contradiction with any known theoretical and experimental results.
\begin{table}[h!]
\centering
\caption{Pad\'e-Borel-Leroy estimate of critical exponents $\gamma$ obtained from six-loop $\sqrt{\varepsilon}$ expansion under averaging through the highest-order and the most stable approximants under the optimal value of the shift parameter $b_{opt} = 7.52$.}
\label{pbl_gamma}
\renewcommand{\tabcolsep}{0.165cm}
\begin{tabular}{{c}|*{6}{c}}
\hline
\hline
$M \setminus L$ & 0 & 1 & 2 & 3 & 4&5 \\
\hline
0&1.    & 1.1682 & 1.1396 & 1.2185 & 1.7830 & 2.2236 \\
1&1.2084 & 1.1441 & 1.1603 & 1.1251 & 1.2497 & \text{} \\
2&1.1217 & 1.1723 & 1.1362 & 1.1805 & \text{} & \text{} \\
3&1.1961 & 1.0897 & 1.2468 & \text{} & \text{} & \text{} \\
4&1.4047 & 1.1754 & \text{} & \text{} & \text{} & \text{} \\
5&0.7371 & \text{} & \text{} & \text{} & \text{} & \text{} \\
\hline
\hline
\end{tabular}
\end{table}
\begin{figure}[h!]
    \includegraphics[width=0.48\textwidth]{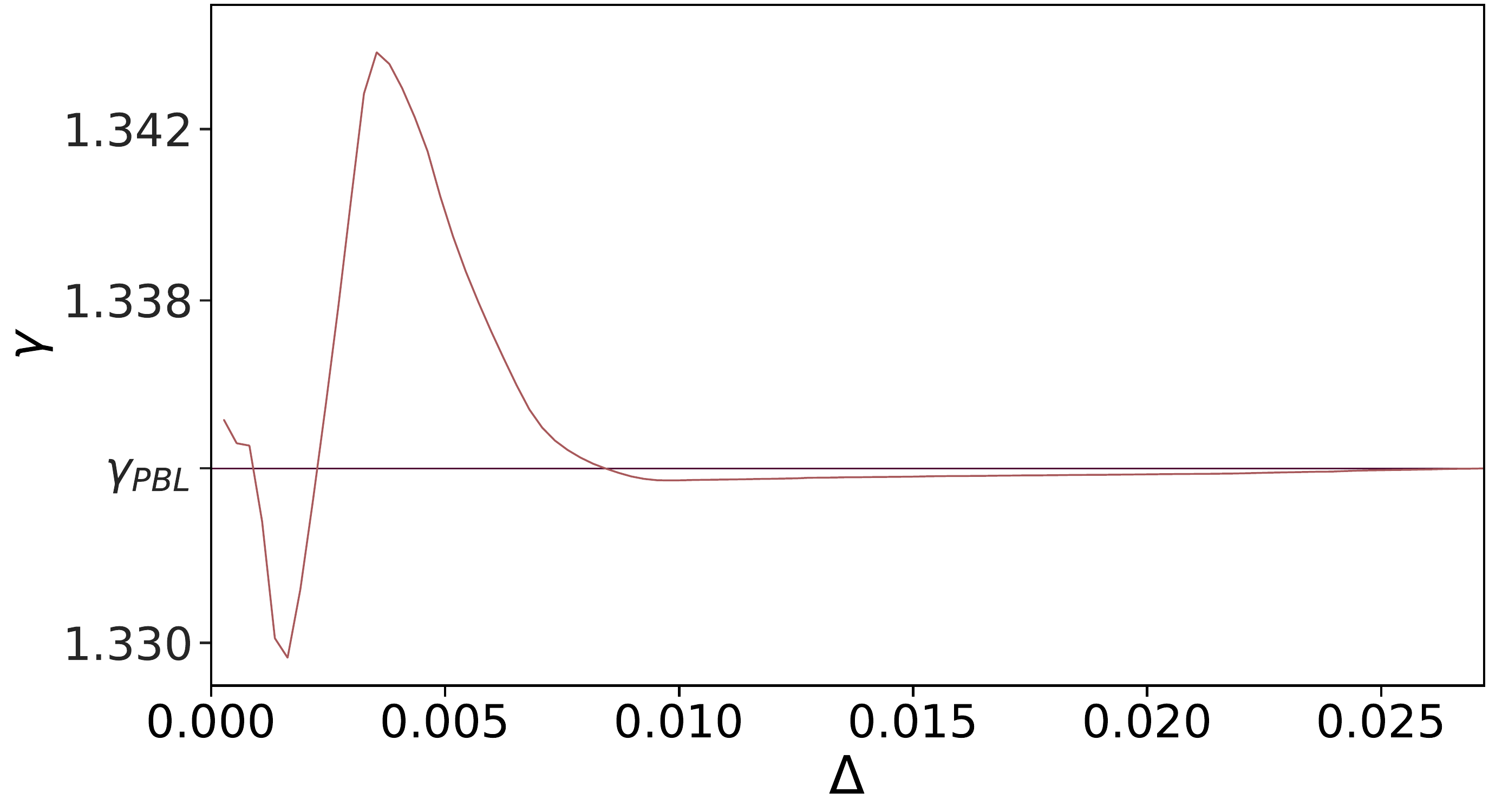}
    \caption{The dependence of critical exponent $\gamma$ estimate on the maximally allowed value of spreading in the analyzed array when PBLPBL technique is applied. The plateau value is $1.3341$.}
    \label{gamma_PBLPBL_spre_fun}
\end{figure}
\begin{figure}[h!]
    \includegraphics[width=0.48\textwidth]{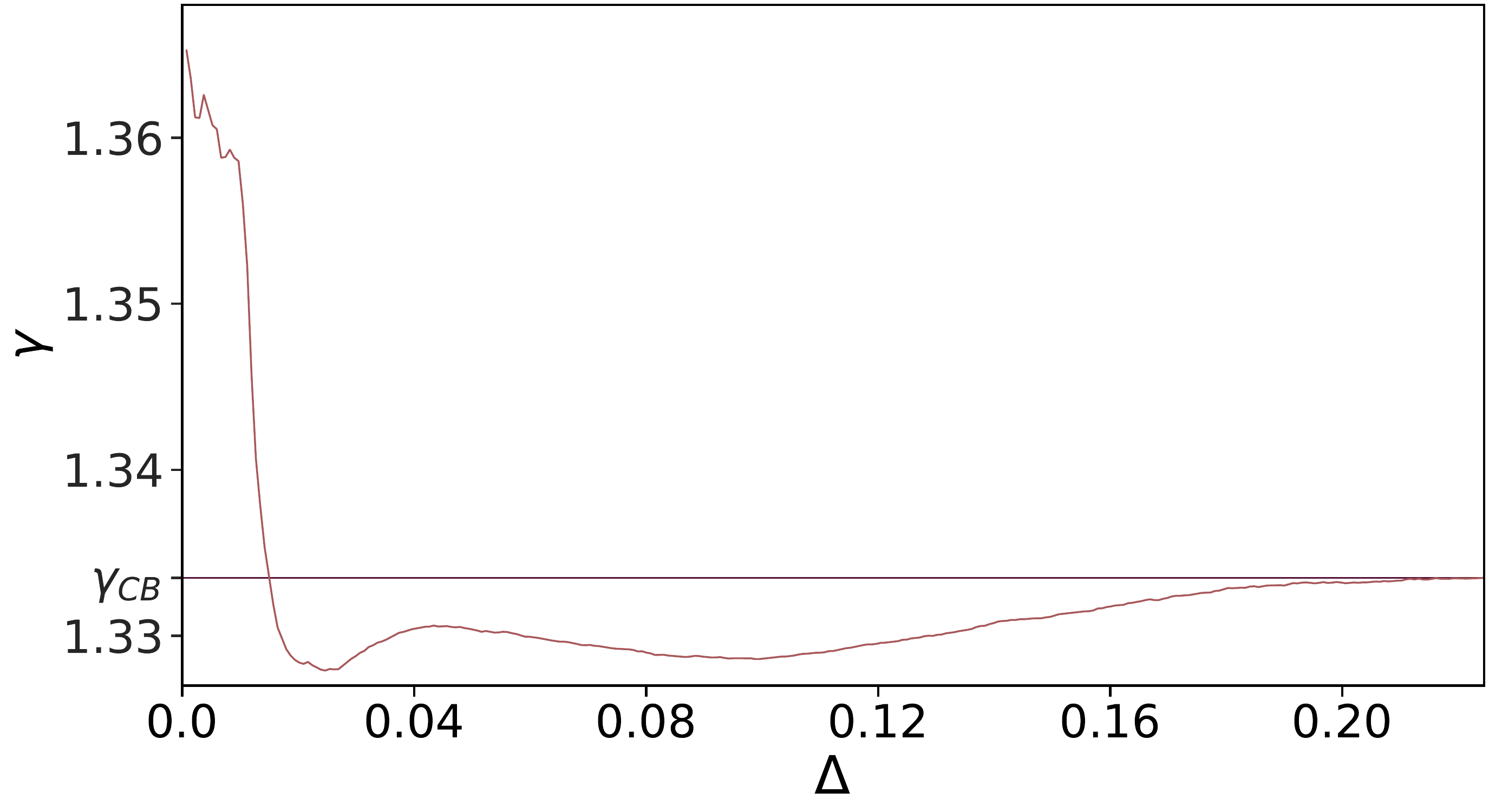}
    \caption{The dependence of critical exponent $\gamma$ estimate on the maximally allowed value of spreading in the analyzed array when CBPBL technique is applied. The plateau value is $1.3335$.}
    \label{gamma_CBPBL_spre_fun}
\end{figure}
\begin{figure}[h!]
    \centering
    \includegraphics[width=0.46\textwidth]{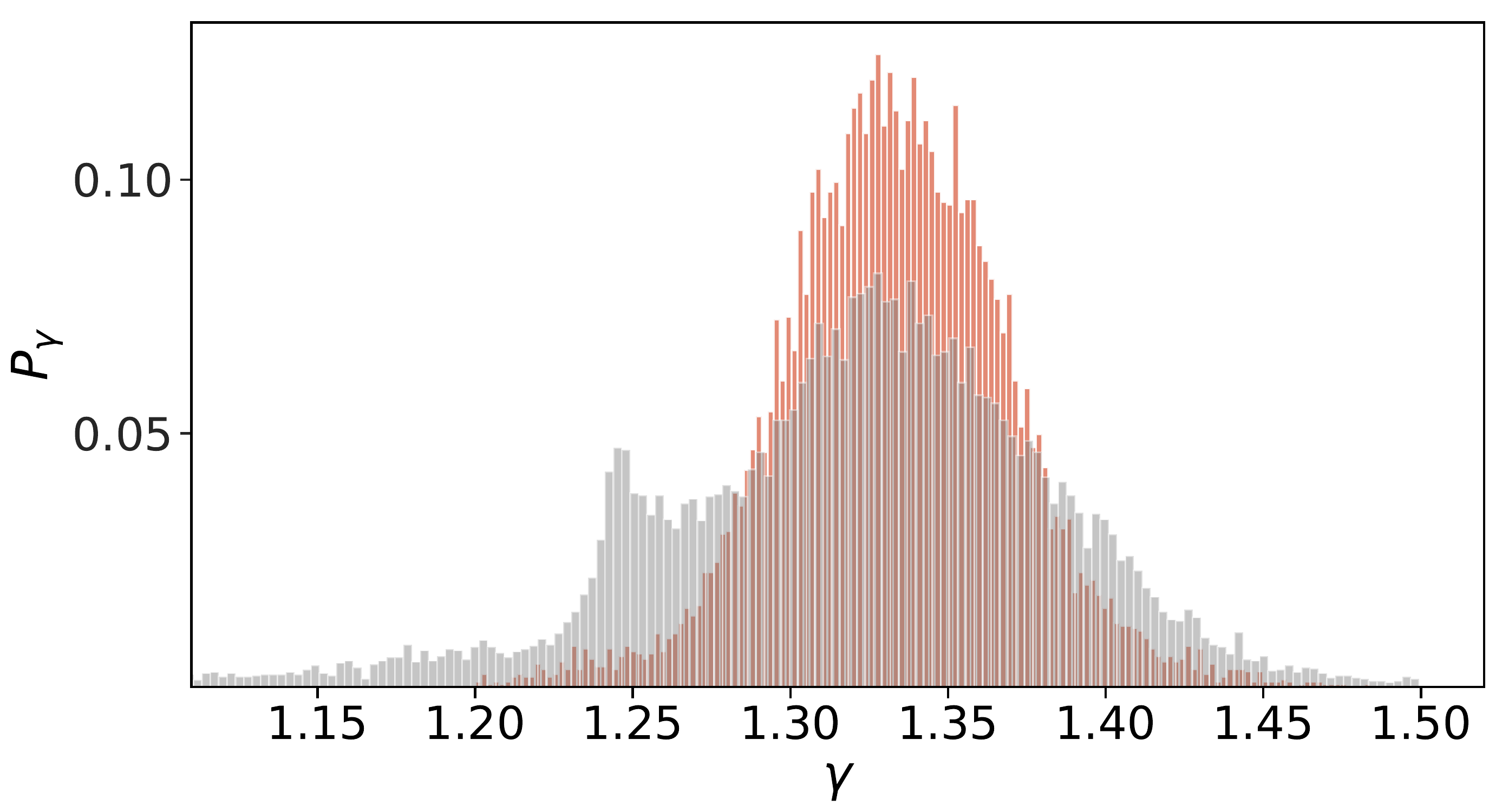}
    \caption{The distribution (red one or more "Gaussian-like") of estimates of critical exponent $\gamma$ when CBPBL technique is applied under  $\Delta=0.22445$. The "gray" distribution corresponds to $\Delta=0.4$.}
    \label{gamma_CBPBL_hist}
\end{figure}

\noindent The transformed RG expansions for $\gamma$ are as follows:
\begin{eqnarray}\label{gamma_rg_exp_num_both}
&&\gamma_{PBL}=1+0.317(14)z-0.0351(99)z^{2}+0.224(42)z^{3}\nonumber\\
&&\ -0.73(17)z^{4}+3.09(92)z^{5}-15.3(5.5)z^{6}+\bigo{z^{7}},\quad\\
&&\gamma_{CB}=1.0+0.290(51)z-0.046(34)z^{2}+ 0.20(14)z^{3}\nonumber\\
&&\ -0.61(54)z^{4}+2.5(2.8)z^{5}-12(16)z^{6}+\bigo{z^{7}}.\quad
\end{eqnarray}
Corresponding numerical estimates obtained within two strategies described are
\begin{eqnarray}
     \gamma_{PBL}=1.334(41),\quad \gamma_{CB}=1.334(36).
\end{eqnarray}
Hence, we admit 
\begin{eqnarray}
     \gamma=1.334(38).
\end{eqnarray}
as the final result of our calculations. Corresponding dependencies of mean value and standard deviation on the maximally allowed value of spreading for PBLPBL and CBPBL cases, along with the distribution of estimates for $\gamma$ the CBPBL machinery yields, are depicted in Fig.~\ref{gamma_PBLPBL_spre_fun}, \ref{gamma_CBPBL_spre_fun}, and 9 respectively.

\subsubsection{Correction-to-scaling exponents $\omega_1$ and $\omega_2$}
Physically important quantities include correction-to-scaling exponents $\omega_1$ and $\omega_2$. Starting from the definition \eqref{stab_matrix} one can find $\sqrt{\epsilon}$ expansions for $\omega_1$ and $\omega_2$. In decimals, they are as follows:
\begin{eqnarray}\label{omega_1_sqrt_num_exp}
&&\omega_1=0.672927\varepsilon^{1/2}-1.92551\varepsilon-0.572525\varepsilon^{3/2}\nonumber\\
&&\qquad-13.9313\varepsilon^{2}-69.3309\varepsilon^{5/2}+\bigo{\varepsilon^{3}},\ \\
\label{omega_2_sqrt_num_exp}
&&\quad\omega_2=2\varepsilon+3.70401\varepsilon^{3/2}+11.3087\varepsilon^{2}\nonumber\\
&&\quad\qquad\qquad+64.072\varepsilon^{5/2}+\bigo{\varepsilon^{3}}.
\end{eqnarray}
\begin{figure}[b!]
    \centering
    \includegraphics[width=0.46\textwidth]{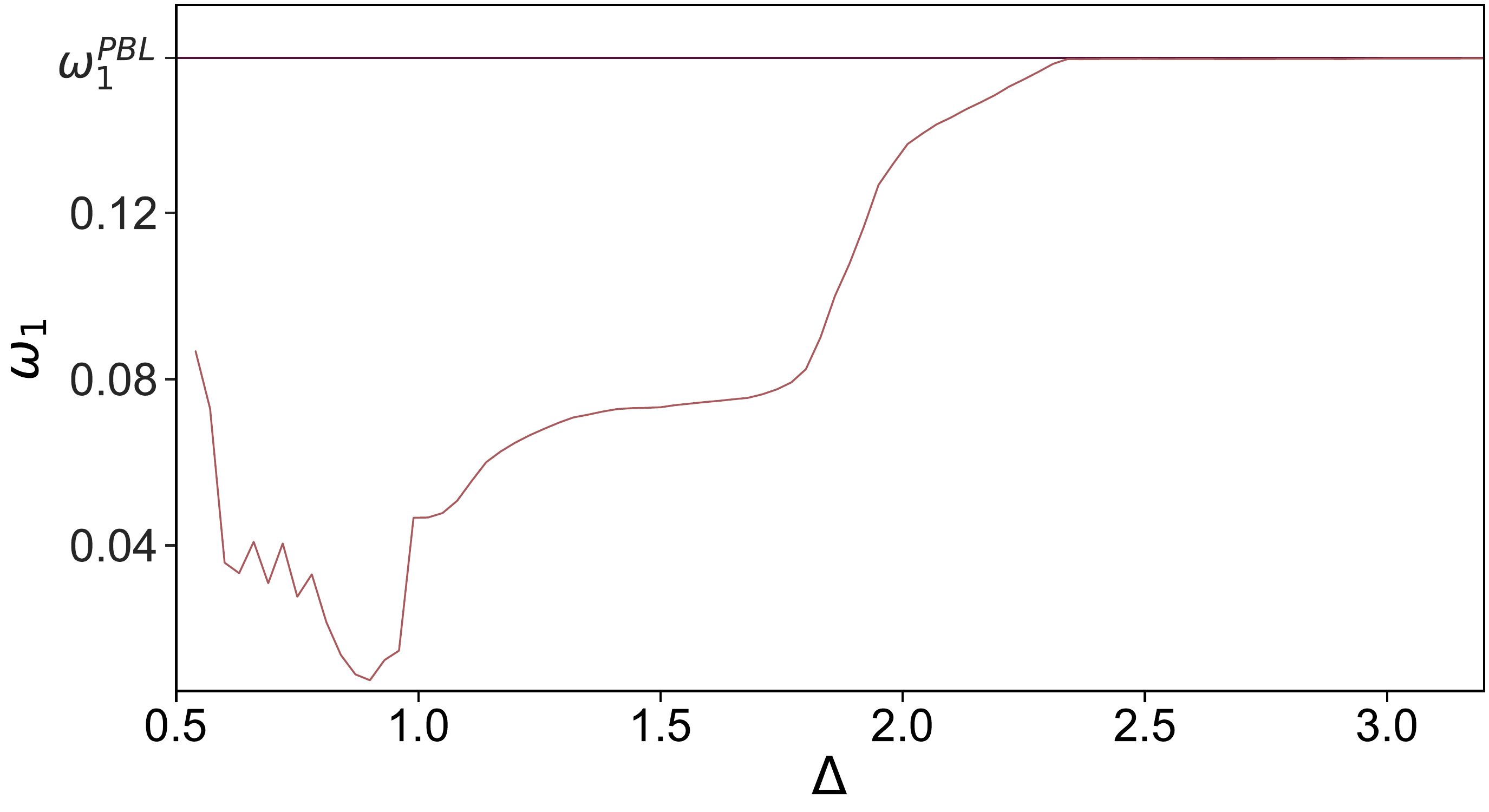}
    \caption{The dependence of $\omega_1$ PBLPBL estimate on maximally allowed spreading of points entering the sample on the basis of which the mentioned estimate is obtained. The plateau value is $0.2496$.}
    \label{omega1_PBLPBL}
\end{figure}
\begin{figure}[b!]
    \centering
    \includegraphics[width=0.46\textwidth]{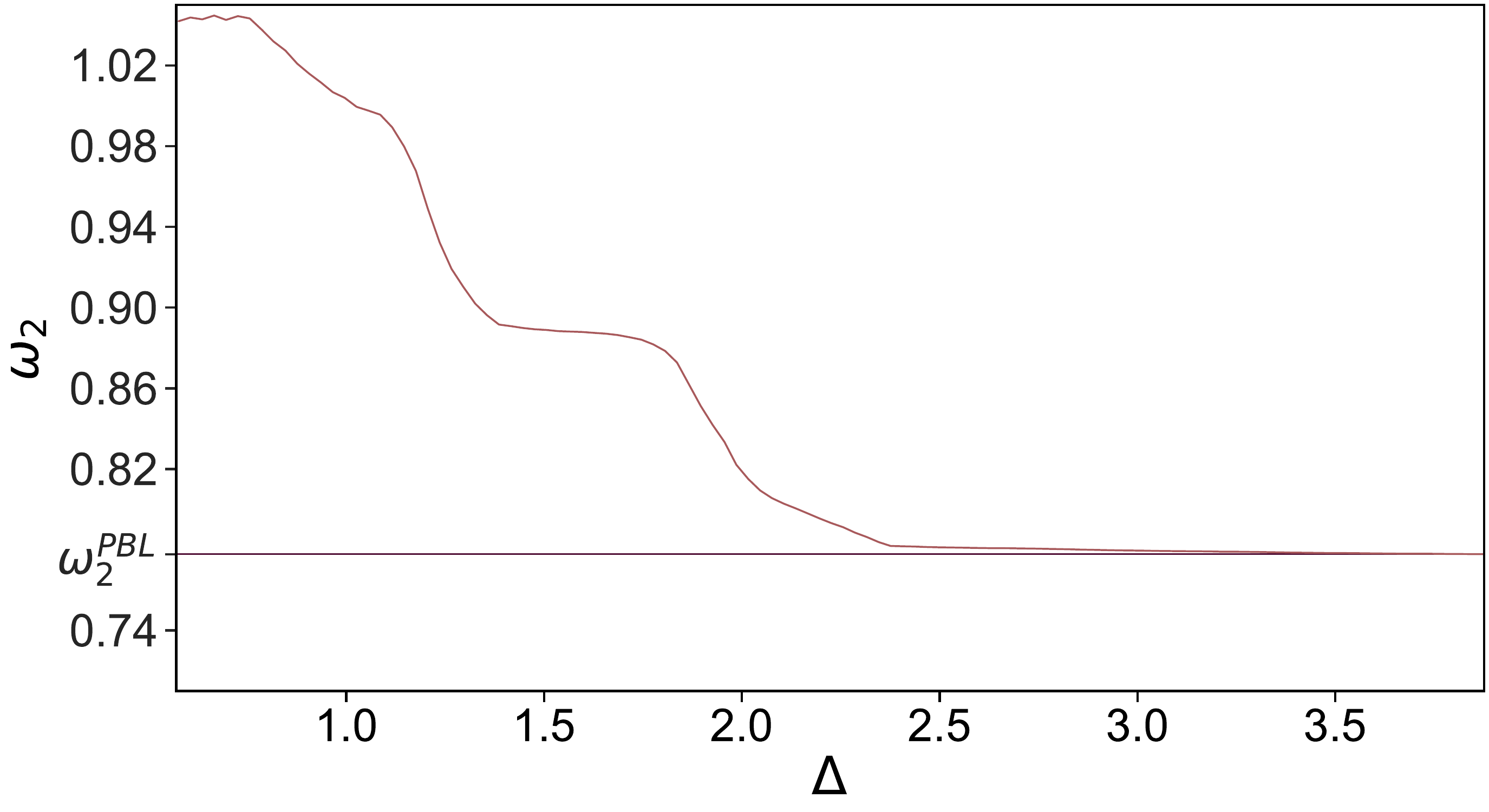}
    \caption{The dependence of $\omega_2$ PBLPBL estimate on maximally allowed spreading of points entering the sample on the basis of which the mentioned estimate is obtained. The plateau value is $0.7778$.}
    \label{omega2_PBLPBL}
\end{figure}

\noindent Resummation of these $\sqrt{\varepsilon}$ expansions is obviously meaningless because of their quite unfavorable structure. Therefore we move to the 3D resummation procedure,  without treatment of $\sqrt{\varepsilon}$ series themselves. Having applied both previously described procedures to $\partial_{g_1}\beta_1$, $\partial_{g_1}\beta_1$, $\partial_{g_2}\beta_2$, and $\partial_{g_2}\beta_2$, we can calculate the eigenvalues of the matrix~\eqref{stab_matrix}. In case of PBLPBL technique we come to the estimates
\begin{eqnarray}
\omega_{1,PBL}=0.15(10), \quad \omega_{2,PBL}=0.78(12).
\end{eqnarray}
Corresponding dependencies on the spreading value for $\omega_1$ and $\omega_2$ are presented in Fig.~\ref{omega1_PBLPBL} and Fig.~\ref{omega2_PBLPBL} respectively. The positiveness of these numbers confirms the conclusion that the random fixed point is stable and governs the critical behavior of 3D random Ising model. The application of CBPBL resummation procedure to the series for $\omega_1$ and $\omega_2$, however, does not allow one to extract any stable estimates.

\subsubsection{Zoo of critical exponents for RIM -- final estimates}
In order to get numerical estimates for the rest of critical exponents -- $\alpha$, $\beta$, $\delta$, and $\eta$  -- we can address well known scaling relations. Corresponding numbers  obtained on the base of the estimates for $\nu$ and $\gamma$ via scaling relations, along with those found in previous subsections, are presented in Table \ref{cr_exp_estimates_this_work}.

\begin{table}[t!]
\begin{center}
\caption{Numerical estimates of critical exponents characterising behavior of three-dimensional random Ising model at criticality. Stars as superscripts denote critical exponents evaluated with a help of scaling relations.}
\label{cr_exp_estimates_this_work}
\setlength{\tabcolsep}{11.8pt}
\begin{tabular}{cccc}
\hline
\hline
$\alpha^*$ & $\beta^*$ & $\gamma$ & $\delta^*$\\
\hline
$-0.025(58)$ &$0.346(34)$ & $1.334(38)$ & $4.86(45)$ \\
\hline
$\eta^*$&  $\nu$ & $\omega_1$&$\omega_2$ \\
\hline
$0.024(79)$& $0.675(19)$&$0.15(10)$&$0.78(12)$ \\
\hline
\hline
\end{tabular}
\end{center}
\end{table}

\section{Conclusion}
To conclude, we have calculated six-loop $\sqrt{\varepsilon}$ expansions for critical exponents of weakly disordered Ising model as well as those for the random fixed point coordinates. Resummation of $\sqrt{\varepsilon}$ expansions themselves by means of techniques using Pad\'e--Borel--Leroy transformation  failed to provide reasonable numerical estimates, confirming the conjecture that such series are not Borel summable. Alternative resummation machinery dealing with the expansions in renormalized couplings at ${\varepsilon}=1$ enabled to find stable numerical estimates for critical exponents and the fixed point location. Numerical values of critical exponents thus obtained turned out to be in a good agreement with their counterparts extracted from computer simulations and experimental data.                 

\begin{acknowledgments}
M.V.K. and A.K. gratefully acknowledge the support of Foundation for the Advancement of Theoretical Physics "BASIS" through Grant 18-1-2-43-1. The work by A.I.S. was supported by Grant of the Russian Science Foundation No 20-11-20226.
\end{acknowledgments}

\appendix*
\section{The information about Supplementary Materials}
The RG expansions for $\beta$-functions and anomalous dimensions in six-loop approximation are presented in Supplementary Materials as \textit{Mathematica}-file (\textit{rg\_expansions\_im\_is.m}). The $\sqrt{\varepsilon}$ expansions for random FP coordinates can be found there as \textit{Mathematica}-file (\textit{fp\_coordinates\_im\_is.m}). The $\sqrt{\varepsilon}$ expansions for critical exponents of RIM universality class are presented in Supplementary Materials as \textit{Mathematica}-file (\textit{critical\_exponents\_im\_is.m}).

\section{The $\sqrt{\varepsilon}$ expansions for $\nu$ and $\gamma$}
Here we present the six-loop $\sqrt{\varepsilon}$ expansions for critical exponents $\nu$ and $\gamma$ with coefficients in analytical form:
\begin{widetext}
\begingroup
\allowdisplaybreaks
\begin{eqnarray}\label{nu_sqrt_epsilon}
\nu&=&\frac{1}{2}+\varepsilon^{1/2}\frac{1}{2}\sqrt{\frac{3}{106}}+\varepsilon \left[\frac{535}{22472}-\frac{189}{5618}\zeta (3)\right]\nonumber\\
&&+\varepsilon^{3/2}\left[\frac{591133}{4764064}\sqrt{\frac{3}{106}}+\frac{39555}{148877}\sqrt{\frac{3}{106}}\zeta (3)-\frac{1215}{2809}\sqrt{\frac{3}{106}}\zeta (5)+\frac{59535}{297754}\sqrt{\frac{3}{106}}\zeta (3)^2\right]\nonumber\\
&&+\varepsilon^{2}\left[\frac{64154587}{6691127888}+\frac{365093775}{6691127888}\zeta (3)-\frac{567}{22472}\zeta (4)-\frac{42465465}{418195493}\zeta (3)^2+\frac{3788415}{63123848}\zeta (5)+\frac{392931}{2382032}\zeta (7)\right.\nonumber\\
&&\left.+\frac{1377810}{7890481}\zeta (3)\zeta (5)-\frac{18003384}{418195493}\zeta (3)^3\right]+\varepsilon^{5/2}\left[\frac{1926534225575}{22696305796096}\sqrt{\frac{3}{106}}+\frac{118665}{595508}\sqrt{\frac{3}{106}}\zeta (4)\right.\nonumber\\
&&\left.-\frac{6075}{11236}\sqrt{\frac{3}{106}}\zeta (6)+\frac{357696}{744385}\sqrt{\frac{6}{53}}\zeta(3,5)-\frac{129611897103}{354629778064}\sqrt{\frac{3}{106}}\zeta (3)-\frac{864432}{744385}\sqrt{\frac{6}{53}}\zeta (8)\right.\nonumber\\
&&+\frac{178605}{595508}\sqrt{\frac{3}{106}}\zeta (4)\zeta (3)+\frac{293024828745}{709259556128}\sqrt{\frac{3}{106}}\zeta (3)^2+\frac{35474659317}{22164361129}\sqrt{\frac{3}{106}}\zeta (3)^3+\frac{9496728819}{6691127888}\sqrt{\frac{3}{106}}\zeta (5)\nonumber\\
&&+\frac{731097981}{1672781972}\sqrt{\frac{3}{106}}\zeta (3)\zeta (5)+\frac{32750405919}{88657444516}\sqrt{\frac{3}{106}}\zeta (3)^4-\frac{401698521}{126247696}\sqrt{\frac{3}{106}}\zeta (7)+\frac{10333575}{7890481}\sqrt{\frac{3}{106}}\zeta (5)^2\nonumber\\
&&-\frac{911421315}{418195493}\sqrt{\frac{3}{106}}\zeta (3)^2\zeta (5)\left.+\frac{456885}{148877}\sqrt{\frac{6}{53}}\zeta (9)-\frac{173282571}{63123848}\sqrt{\frac{3}{106}}\zeta (3)\zeta (7)\right]+\bigo{\varepsilon^{3}},\\
\label{gamma_sqrt_full}
\gamma&=&1+\varepsilon^{1/2}\sqrt{\frac{3}{106}}+\varepsilon \left[\frac{147}{2809}-\frac{189}{2809}\zeta (3)\right]\nonumber\\
&&+\varepsilon^{3/2}\left[\frac{419201}{2382032}\sqrt{\frac{3}{106}}+\frac{75771}{148877}\sqrt{\frac{3}{106}}\zeta (3)-\frac{1215}{2809}\sqrt{\frac{6}{53}}\zeta (5)+\frac{59535}{148877}\sqrt{\frac{3}{106}}\zeta (3)^2\right]\nonumber\\
&&+\varepsilon^{2}\left[\frac{18425409}{836390986}+\frac{416314989}{3345563944}\zeta (3)-\frac{567}{11236}\zeta (4)-\frac{83037717}{418195493}\zeta (3)^2+\frac{3530835}{31561924}\zeta (5)+\frac{392931}{1191016}\zeta (7)\right.\nonumber\\
&&\left.+\frac{2755620}{7890481}\zeta (3)\zeta (5)-\frac{36006768}{418195493}\zeta (3)^3\right]+\varepsilon^{5/2}\left[\frac{1063453114159}{11348152898048}\sqrt{\frac{3}{106}}+\frac{227313}{595508}\sqrt{\frac{3}{106}}\zeta (4)-\frac{6075}{5618}\sqrt{\frac{3}{106}}\zeta (6)\right.\nonumber\\
&&+\frac{715392}{744385}\sqrt{\frac{6}{53}}\zeta(3,5)-\frac{288766951863}{354629778064}\sqrt{\frac{3}{106}}\zeta (3)-\frac{1728864}{744385}\sqrt{\frac{6}{53}}\zeta (8)+\frac{178605}{297754}\sqrt{\frac{3}{106}}\zeta (4)\zeta (3)\nonumber\\
&&+\frac{236207403045}{354629778064}\sqrt{\frac{3}{106}}\zeta (3)^2+\frac{70114411701}{22164361129}\sqrt{\frac{3}{106}}\zeta (3)^3+\frac{10063304649}{3345563944}\sqrt{\frac{3}{106}}\zeta (5)+\frac{844690761}{836390986}\sqrt{\frac{3}{106}}\zeta (3)\zeta (5)\nonumber\\
&&+\frac{32750405919}{44328722258}\sqrt{\frac{3}{106}}\zeta (3)^4-\frac{98689185}{15780962}\sqrt{\frac{3}{106}}\zeta (7)+\frac{10333575}{7890481}\sqrt{\frac{6}{53}}\zeta (5)^2-\frac{911421315}{418195493}\sqrt{\frac{6}{53}}\zeta (3)^2\zeta (5)\nonumber\\
&&\left.+\frac{913770}{148877}\sqrt{\frac{6}{53}}\zeta (9)-\frac{173282571}{31561924}\sqrt{\frac{3}{106}}\zeta (3)\zeta (7)\right]+\bigo{\varepsilon^{3}}.
\end{eqnarray}
\endgroup
\end{widetext}
\bibliography{ImpureIsing}

\begin{thebibliography}{57}%
\makeatletter
\providecommand \@ifxundefined [1]{%
 \@ifx{#1\undefined}
}%
\providecommand \@ifnum [1]{%
 \ifnum #1\expandafter \@firstoftwo
 \else \expandafter \@secondoftwo
 \fi
}%
\providecommand \@ifx [1]{%
 \ifx #1\expandafter \@firstoftwo
 \else \expandafter \@secondoftwo
 \fi
}%
\providecommand \natexlab [1]{#1}%
\providecommand \enquote  [1]{``#1''}%
\providecommand \bibnamefont  [1]{#1}%
\providecommand \bibfnamefont [1]{#1}%
\providecommand \citenamefont [1]{#1}%
\providecommand \href@noop [0]{\@secondoftwo}%
\providecommand \href [0]{\begingroup \@sanitize@url \@href}%
\providecommand \@href[1]{\@@startlink{#1}\@@href}%
\providecommand \@@href[1]{\endgroup#1\@@endlink}%
\providecommand \@sanitize@url [0]{\catcode `\\12\catcode `\$12\catcode
  `\&12\catcode `\#12\catcode `\^12\catcode `\_12\catcode `\%12\relax}%
\providecommand \@@startlink[1]{}%
\providecommand \@@endlink[0]{}%
\providecommand \url  [0]{\begingroup\@sanitize@url \@url }%
\providecommand \@url [1]{\endgroup\@href {#1}{\urlprefix }}%
\providecommand \urlprefix  [0]{URL }%
\providecommand \Eprint [0]{\href }%
\providecommand \doibase [0]{http://dx.doi.org/}%
\providecommand \selectlanguage [0]{\@gobble}%
\providecommand \bibinfo  [0]{\@secondoftwo}%
\providecommand \bibfield  [0]{\@secondoftwo}%
\providecommand \translation [1]{[#1]}%
\providecommand \BibitemOpen [0]{}%
\providecommand \bibitemStop [0]{}%
\providecommand \bibitemNoStop [0]{.\EOS\space}%
\providecommand \EOS [0]{\spacefactor3000\relax}%
\providecommand \BibitemShut  [1]{\csname bibitem#1\endcsname}%
\let\auto@bib@innerbib\@empty
\bibitem [{\citenamefont {Harris}\ and\ \citenamefont
  {Lubensky}(1974)}]{PhysRevLett.33.1540}%
  \BibitemOpen
  \bibfield  {author} {\bibinfo {author} {\bibfnamefont {A.~B.}\ \bibnamefont
  {Harris}}\ and\ \bibinfo {author} {\bibfnamefont {T.~C.}\ \bibnamefont
  {Lubensky}},\ }\href@noop {} {\bibfield  {journal} {\bibinfo  {journal}
  {Phys. Rev. Lett.}\ }\textbf {\bibinfo {volume} {33}},\ \bibinfo {pages}
  {1540} (\bibinfo {year} {1974})}\BibitemShut {NoStop}%
\bibitem [{\citenamefont {Lubensky}(1975)}]{PhysRevB.11.3573}%
  \BibitemOpen
  \bibfield  {author} {\bibinfo {author} {\bibfnamefont {T.~C.}\ \bibnamefont
  {Lubensky}},\ }\href@noop {} {\bibfield  {journal} {\bibinfo  {journal}
  {Phys. Rev. B}\ }\textbf {\bibinfo {volume} {11}},\ \bibinfo {pages} {3573}
  (\bibinfo {year} {1975})}\BibitemShut {NoStop}%
\bibitem [{\citenamefont {Khmelnitskii}(1975)}]{Khmelnitskii1975}%
  \BibitemOpen
  \bibfield  {author} {\bibinfo {author} {\bibfnamefont {D.~E.}\ \bibnamefont
  {Khmelnitskii}},\ }\href@noop {} {\bibfield  {journal} {\bibinfo  {journal}
  {Zh. Eksp. Teor. Fiz.}\ }\textbf {\bibinfo {volume} {68}},\ \bibinfo {pages}
  {1960} (\bibinfo {year} {1975})},\ \bibinfo {note} {[Sov. Phys. JETP 41, 981
  (1975)]}\BibitemShut {NoStop}%
\bibitem [{\citenamefont {Grinstein}\ and\ \citenamefont
  {Luther}(1976)}]{PhysRevB.13.1329}%
  \BibitemOpen
  \bibfield  {author} {\bibinfo {author} {\bibfnamefont {G.}~\bibnamefont
  {Grinstein}}\ and\ \bibinfo {author} {\bibfnamefont {A.}~\bibnamefont
  {Luther}},\ }\href@noop {} {\bibfield  {journal} {\bibinfo  {journal} {Phys.
  Rev. B}\ }\textbf {\bibinfo {volume} {13}},\ \bibinfo {pages} {1329}
  (\bibinfo {year} {1976})}\BibitemShut {NoStop}%
\bibitem [{\citenamefont {Harris}(1974)}]{Harris_1974}%
  \BibitemOpen
  \bibfield  {author} {\bibinfo {author} {\bibfnamefont {A.~B.}\ \bibnamefont
  {Harris}},\ }\href@noop {} {\bibfield  {journal} {\bibinfo  {journal} {J.
  Phys. C}\ }\textbf {\bibinfo {volume} {7}},\ \bibinfo {pages} {1671}
  (\bibinfo {year} {1974})}\BibitemShut {NoStop}%
\bibitem [{\citenamefont {Brody}\ and\ \citenamefont {Meier}(2015)}]{BM2015}%
  \BibitemOpen
  \bibfield  {author} {\bibinfo {author} {\bibfnamefont {D.~C.}\ \bibnamefont
  {Brody}}\ and\ \bibinfo {author} {\bibfnamefont {D.~M.}\ \bibnamefont
  {Meier}},\ }\href@noop {} {\bibfield  {journal} {\bibinfo  {journal} {Phys.
  Rev. Lett.}\ }\textbf {\bibinfo {volume} {114}},\ \bibinfo {pages} {100502}
  (\bibinfo {year} {2015})}\BibitemShut {NoStop}%
\bibitem [{\citenamefont {Brody}\ \emph {et~al.}(2015)\citenamefont {Brody},
  \citenamefont {Gibbons},\ and\ \citenamefont {Meier}}]{BGM15}%
  \BibitemOpen
  \bibfield  {author} {\bibinfo {author} {\bibfnamefont {D.~C.}\ \bibnamefont
  {Brody}}, \bibinfo {author} {\bibfnamefont {G.~W.}\ \bibnamefont {Gibbons}},
  \ and\ \bibinfo {author} {\bibfnamefont {D.~M.}\ \bibnamefont {Meier}},\
  }\href@noop {} {\bibfield  {journal} {\bibinfo  {journal} {New J. Phys.}\
  }\textbf {\bibinfo {volume} {17}},\ \bibinfo {pages} {033048} (\bibinfo
  {year} {2015})}\BibitemShut {NoStop}%
\bibitem [{\citenamefont {Aharony}(1976)}]{AA76}%
  \BibitemOpen
  \bibfield  {author} {\bibinfo {author} {\bibfnamefont {A.}~\bibnamefont
  {Aharony}},\ }\href@noop {} {\bibfield  {journal} {\bibinfo  {journal} {Phys.
  Rev. B}\ }\textbf {\bibinfo {volume} {13}},\ \bibinfo {pages} {2092}
  (\bibinfo {year} {1976})}\BibitemShut {NoStop}%
\bibitem [{\citenamefont {Jayaprakash}\ and\ \citenamefont
  {Katz}(1977)}]{PhysRevB.16.3987}%
  \BibitemOpen
  \bibfield  {author} {\bibinfo {author} {\bibfnamefont {C.}~\bibnamefont
  {Jayaprakash}}\ and\ \bibinfo {author} {\bibfnamefont {H.~J.}\ \bibnamefont
  {Katz}},\ }\href@noop {} {\bibfield  {journal} {\bibinfo  {journal} {Phys.
  Rev. B}\ }\textbf {\bibinfo {volume} {16}},\ \bibinfo {pages} {3987}
  (\bibinfo {year} {1977})}\BibitemShut {NoStop}%
\bibitem [{\citenamefont {Shalaev}(1977)}]{SHALAEV1977}%
  \BibitemOpen
  \bibfield  {author} {\bibinfo {author} {\bibfnamefont {B.~N.}\ \bibnamefont
  {Shalaev}},\ }\href@noop {} {\bibfield  {journal} {\bibinfo  {journal} {Zh.
  Eksp. Teor. Fiz.}\ }\textbf {\bibinfo {volume} {73}},\ \bibinfo {pages}
  {2301} (\bibinfo {year} {1977})},\ \bibinfo {note} {[Sov. Phys. JETP 46, 1204
  (1977)]}\BibitemShut {NoStop}%
\bibitem [{\citenamefont {Shalaev}\ \emph {et~al.}(1997)\citenamefont
  {Shalaev}, \citenamefont {Antonenko},\ and\ \citenamefont
  {Sokolov}}]{SHALAEV1997105}%
  \BibitemOpen
  \bibfield  {author} {\bibinfo {author} {\bibfnamefont {B.~N.}\ \bibnamefont
  {Shalaev}}, \bibinfo {author} {\bibfnamefont {S.~A.}\ \bibnamefont
  {Antonenko}}, \ and\ \bibinfo {author} {\bibfnamefont {A.~I.}\ \bibnamefont
  {Sokolov}},\ }\href@noop {} {\bibfield  {journal} {\bibinfo  {journal} {Phys.
  Lett. A}\ }\textbf {\bibinfo {volume} {230}},\ \bibinfo {pages} {105}
  (\bibinfo {year} {1997})}\BibitemShut {NoStop}%
\bibitem [{\citenamefont {Folk}\ \emph {et~al.}(1999)\citenamefont {Folk},
  \citenamefont {Holovatch},\ and\ \citenamefont
  {Yavors'kii}}]{FOLKHOLYAR1999}%
  \BibitemOpen
  \bibfield  {author} {\bibinfo {author} {\bibfnamefont {R.}~\bibnamefont
  {Folk}}, \bibinfo {author} {\bibfnamefont {Y.}~\bibnamefont {Holovatch}}, \
  and\ \bibinfo {author} {\bibfnamefont {T.}~\bibnamefont {Yavors'kii}},\
  }\href@noop {} {\bibfield  {journal} {\bibinfo  {journal} {JETP Lett.}\
  }\textbf {\bibinfo {volume} {69}},\ \bibinfo {pages} {747} (\bibinfo {year}
  {1999})}\BibitemShut {NoStop}%
\bibitem [{\citenamefont {Janssen}\ \emph {et~al.}(1995)\citenamefont
  {Janssen}, \citenamefont {Oerding},\ and\ \citenamefont
  {Sengespeick}}]{Janssen_1995}%
  \BibitemOpen
  \bibfield  {author} {\bibinfo {author} {\bibfnamefont {H.~K.}\ \bibnamefont
  {Janssen}}, \bibinfo {author} {\bibfnamefont {K.}~\bibnamefont {Oerding}}, \
  and\ \bibinfo {author} {\bibfnamefont {E.}~\bibnamefont {Sengespeick}},\
  }\href@noop {} {\bibfield  {journal} {\bibinfo  {journal} {J. Phys. A}\
  }\textbf {\bibinfo {volume} {28}},\ \bibinfo {pages} {6073} (\bibinfo {year}
  {1995})}\BibitemShut {NoStop}%
\bibitem [{\citenamefont {Folk}\ \emph {et~al.}(2000)\citenamefont {Folk},
  \citenamefont {Holovatch},\ and\ \citenamefont
  {Yavors'kii}}]{PhysRevB.61.15114}%
  \BibitemOpen
  \bibfield  {author} {\bibinfo {author} {\bibfnamefont {R.}~\bibnamefont
  {Folk}}, \bibinfo {author} {\bibfnamefont {Y.}~\bibnamefont {Holovatch}}, \
  and\ \bibinfo {author} {\bibfnamefont {T.}~\bibnamefont {Yavors'kii}},\
  }\href@noop {} {\bibfield  {journal} {\bibinfo  {journal} {Phys. Rev. B}\
  }\textbf {\bibinfo {volume} {61}},\ \bibinfo {pages} {15114} (\bibinfo {year}
  {2000})}\BibitemShut {NoStop}%
\bibitem [{\citenamefont {Blavats'ka}\ and\ \citenamefont
  {Holovatch}(2003)}]{BLAVATSKA2003221}%
  \BibitemOpen
  \bibfield  {author} {\bibinfo {author} {\bibfnamefont {V.}~\bibnamefont
  {Blavats'ka}}\ and\ \bibinfo {author} {\bibfnamefont {Y.}~\bibnamefont
  {Holovatch}},\ }\href@noop {} {\bibfield  {journal} {\bibinfo  {journal} {J.
  Mol. Liq.}\ }\textbf {\bibinfo {volume} {105}},\ \bibinfo {pages} {221}
  (\bibinfo {year} {2003})}\BibitemShut {NoStop}%
\bibitem [{\citenamefont {Schloms}\ and\ \citenamefont
  {Dohm}(1987)}]{Schloms_1987}%
  \BibitemOpen
  \bibfield  {author} {\bibinfo {author} {\bibfnamefont {R.}~\bibnamefont
  {Schloms}}\ and\ \bibinfo {author} {\bibfnamefont {V.}~\bibnamefont {Dohm}},\
  }\href@noop {} {\bibfield  {journal} {\bibinfo  {journal} {Eur. Phys. Lett.}\
  }\textbf {\bibinfo {volume} {3}},\ \bibinfo {pages} {413} (\bibinfo {year}
  {1987})}\BibitemShut {NoStop}%
\bibitem [{\citenamefont {Schloms}\ and\ \citenamefont
  {Dohm}(1989)}]{SCHLOMS1989639}%
  \BibitemOpen
  \bibfield  {author} {\bibinfo {author} {\bibfnamefont {R.}~\bibnamefont
  {Schloms}}\ and\ \bibinfo {author} {\bibfnamefont {V.}~\bibnamefont {Dohm}},\
  }\href@noop {} {\bibfield  {journal} {\bibinfo  {journal} {Nucl. Phys. B}\
  }\textbf {\bibinfo {volume} {328}},\ \bibinfo {pages} {639} (\bibinfo {year}
  {1989})}\BibitemShut {NoStop}%
\bibitem [{\citenamefont {Jug}(1983)}]{JG83}%
  \BibitemOpen
  \bibfield  {author} {\bibinfo {author} {\bibfnamefont {G.}~\bibnamefont
  {Jug}},\ }\href@noop {} {\bibfield  {journal} {\bibinfo  {journal} {Phys.
  Rev. B}\ }\textbf {\bibinfo {volume} {27}},\ \bibinfo {pages} {609} (\bibinfo
  {year} {1983})}\BibitemShut {NoStop}%
\bibitem [{\citenamefont {Holovatch}\ and\ \citenamefont
  {Shpot}(1992)}]{Holovatch1992}%
  \BibitemOpen
  \bibfield  {author} {\bibinfo {author} {\bibfnamefont {Y.}~\bibnamefont
  {Holovatch}}\ and\ \bibinfo {author} {\bibfnamefont {M.}~\bibnamefont
  {Shpot}},\ }\href@noop {} {\bibfield  {journal} {\bibinfo  {journal} {J.
  Stat. Phys.}\ }\textbf {\bibinfo {volume} {66}},\ \bibinfo {pages} {867}
  (\bibinfo {year} {1992})}\BibitemShut {NoStop}%
\bibitem [{\citenamefont {Sokolov}\ and\ \citenamefont {Shalaev}(1981)}]{SS81}%
  \BibitemOpen
  \bibfield  {author} {\bibinfo {author} {\bibfnamefont {A.~I.}\ \bibnamefont
  {Sokolov}}\ and\ \bibinfo {author} {\bibfnamefont {B.~N.}\ \bibnamefont
  {Shalaev}},\ }\href@noop {} {\bibfield  {journal} {\bibinfo  {journal} {Fiz.
  Tverd. Tela}\ }\textbf {\bibinfo {volume} {23}},\ \bibinfo {pages} {2058}
  (\bibinfo {year} {1981})},\ \bibinfo {note} {[Sov. Phys. Solid State 23, 1200
  (1981)]}\BibitemShut {NoStop}%
\bibitem [{\citenamefont {Maier}\ and\ \citenamefont
  {Sokolov}(1984)}]{maiersokolov3loop3d}%
  \BibitemOpen
  \bibfield  {author} {\bibinfo {author} {\bibfnamefont {I.~O.}\ \bibnamefont
  {Maier}}\ and\ \bibinfo {author} {\bibfnamefont {A.~I.}\ \bibnamefont
  {Sokolov}},\ }\href@noop {} {\bibfield  {journal} {\bibinfo  {journal} {Fiz.
  Tverd. Tela}\ }\textbf {\bibinfo {volume} {26}},\ \bibinfo {pages} {3454}
  (\bibinfo {year} {1984})},\ \bibinfo {note} {[Sov. Phys. Solid State 26, 2076
  (1984)]}\BibitemShut {NoStop}%
\bibitem [{\citenamefont {Shpot}(1989)}]{SHPOT1989474}%
  \BibitemOpen
  \bibfield  {author} {\bibinfo {author} {\bibfnamefont {N.~A.}\ \bibnamefont
  {Shpot}},\ }\href@noop {} {\bibfield  {journal} {\bibinfo  {journal} {Phys.
  Lett. A}\ }\textbf {\bibinfo {volume} {142}},\ \bibinfo {pages} {474}
  (\bibinfo {year} {1989})}\BibitemShut {NoStop}%
\bibitem [{\citenamefont {Mayer}\ \emph {et~al.}(1989)\citenamefont {Mayer},
  \citenamefont {Sokolov},\ and\ \citenamefont {Shalayev}}]{MSS89}%
  \BibitemOpen
  \bibfield  {author} {\bibinfo {author} {\bibfnamefont {I.~O.}\ \bibnamefont
  {Mayer}}, \bibinfo {author} {\bibfnamefont {A.~I.}\ \bibnamefont {Sokolov}},
  \ and\ \bibinfo {author} {\bibfnamefont {B.~N.}\ \bibnamefont {Shalayev}},\
  }\href@noop {} {\bibfield  {journal} {\bibinfo  {journal} {Ferroelectrics}\
  }\textbf {\bibinfo {volume} {95}},\ \bibinfo {pages} {93} (\bibinfo {year}
  {1989})}\BibitemShut {NoStop}%
\bibitem [{\citenamefont {Mayer}(1989)}]{Mayer_1989}%
  \BibitemOpen
  \bibfield  {author} {\bibinfo {author} {\bibfnamefont {I.~O.}\ \bibnamefont
  {Mayer}},\ }\href@noop {} {\bibfield  {journal} {\bibinfo  {journal} {J.
  Phys. A}\ }\textbf {\bibinfo {volume} {22}},\ \bibinfo {pages} {2815}
  (\bibinfo {year} {1989})}\BibitemShut {NoStop}%
\bibitem [{\citenamefont {Pakhnin}\ and\ \citenamefont {Sokolov}(2000)}]{PS00}%
  \BibitemOpen
  \bibfield  {author} {\bibinfo {author} {\bibfnamefont {D.~V.}\ \bibnamefont
  {Pakhnin}}\ and\ \bibinfo {author} {\bibfnamefont {A.~I.}\ \bibnamefont
  {Sokolov}},\ }\href@noop {} {\bibfield  {journal} {\bibinfo  {journal} {Phys.
  Rev. B}\ }\textbf {\bibinfo {volume} {61}},\ \bibinfo {pages} {15130}
  (\bibinfo {year} {2000})}\BibitemShut {NoStop}%
\bibitem [{\citenamefont {Pelissetto}\ and\ \citenamefont
  {Vicari}(2000)}]{PV00}%
  \BibitemOpen
  \bibfield  {author} {\bibinfo {author} {\bibfnamefont {A.}~\bibnamefont
  {Pelissetto}}\ and\ \bibinfo {author} {\bibfnamefont {E.}~\bibnamefont
  {Vicari}},\ }\href@noop {} {\bibfield  {journal} {\bibinfo  {journal} {Phys.
  Rev. B}\ }\textbf {\bibinfo {volume} {62}},\ \bibinfo {pages} {6393}
  (\bibinfo {year} {2000})}\BibitemShut {NoStop}%
\bibitem [{\citenamefont {Bray}\ \emph {et~al.}(1987)\citenamefont {Bray},
  \citenamefont {McCarthy}, \citenamefont {Moore}, \citenamefont {Reger},\ and\
  \citenamefont {Young}}]{PhysRevB.36.2212}%
  \BibitemOpen
  \bibfield  {author} {\bibinfo {author} {\bibfnamefont {A.~J.}\ \bibnamefont
  {Bray}}, \bibinfo {author} {\bibfnamefont {T.}~\bibnamefont {McCarthy}},
  \bibinfo {author} {\bibfnamefont {M.~A.}\ \bibnamefont {Moore}}, \bibinfo
  {author} {\bibfnamefont {J.~D.}\ \bibnamefont {Reger}}, \ and\ \bibinfo
  {author} {\bibfnamefont {A.~P.}\ \bibnamefont {Young}},\ }\href@noop {}
  {\bibfield  {journal} {\bibinfo  {journal} {Phys. Rev. B}\ }\textbf {\bibinfo
  {volume} {36}},\ \bibinfo {pages} {2212} (\bibinfo {year}
  {1987})}\BibitemShut {NoStop}%
\bibitem [{\citenamefont {McKane}(1994)}]{PhysRevB.49.12003}%
  \BibitemOpen
  \bibfield  {author} {\bibinfo {author} {\bibfnamefont {A.~J.}\ \bibnamefont
  {McKane}},\ }\href@noop {} {\bibfield  {journal} {\bibinfo  {journal} {Phys.
  Rev. B}\ }\textbf {\bibinfo {volume} {49}},\ \bibinfo {pages} {12003}
  (\bibinfo {year} {1994})}\BibitemShut {NoStop}%
\bibitem [{\citenamefont {{\'{A}}lvarez}\ \emph {et~al.}(2000)\citenamefont
  {{\'{A}}lvarez}, \citenamefont {Mart{\'{\i}}n-Mayor},\ and\ \citenamefont
  {Ruiz-Lorenzo}}]{AMR00}%
  \BibitemOpen
  \bibfield  {author} {\bibinfo {author} {\bibfnamefont {G.}~\bibnamefont
  {{\'{A}}lvarez}}, \bibinfo {author} {\bibfnamefont {V.}~\bibnamefont
  {Mart{\'{\i}}n-Mayor}}, \ and\ \bibinfo {author} {\bibfnamefont {J.~J.}\
  \bibnamefont {Ruiz-Lorenzo}},\ }\href@noop {} {\bibfield  {journal} {\bibinfo
   {journal} {J. Phys. A}\ }\textbf {\bibinfo {volume} {33}},\ \bibinfo {pages}
  {841} (\bibinfo {year} {2000})}\BibitemShut {NoStop}%
\bibitem [{\citenamefont {Ballesteros}\ \emph {et~al.}(1998)\citenamefont
  {Ballesteros}, \citenamefont {Fern{\'a}ndez}, \citenamefont
  {Mart{\'\i}n-Mayor}, \citenamefont {Sudupe}, \citenamefont {Parisi},\ and\
  \citenamefont {Ruiz-Lorenzo}}]{BFM98}%
  \BibitemOpen
  \bibfield  {author} {\bibinfo {author} {\bibfnamefont {H.~G.}\ \bibnamefont
  {Ballesteros}}, \bibinfo {author} {\bibfnamefont {L.~A.}\ \bibnamefont
  {Fern{\'a}ndez}}, \bibinfo {author} {\bibfnamefont {V.}~\bibnamefont
  {Mart{\'\i}n-Mayor}}, \bibinfo {author} {\bibfnamefont {A.~M.}\ \bibnamefont
  {Sudupe}}, \bibinfo {author} {\bibfnamefont {G.}~\bibnamefont {Parisi}}, \
  and\ \bibinfo {author} {\bibfnamefont {J.~J.}\ \bibnamefont {Ruiz-Lorenzo}},\
  }\href@noop {} {\bibfield  {journal} {\bibinfo  {journal} {Phys. Rev. B}\
  }\textbf {\bibinfo {volume} {58}},\ \bibinfo {pages} {2740} (\bibinfo {year}
  {1998})}\BibitemShut {NoStop}%
\bibitem [{\citenamefont {Calabrese}\ \emph {et~al.}(2003)\citenamefont
  {Calabrese}, \citenamefont {Martin-Mayor}, \citenamefont {Pelissetto},\ and\
  \citenamefont {Vicari}}]{CMPV03}%
  \BibitemOpen
  \bibfield  {author} {\bibinfo {author} {\bibfnamefont {P.}~\bibnamefont
  {Calabrese}}, \bibinfo {author} {\bibfnamefont {V.}~\bibnamefont
  {Martin-Mayor}}, \bibinfo {author} {\bibfnamefont {A.}~\bibnamefont
  {Pelissetto}}, \ and\ \bibinfo {author} {\bibfnamefont {E.}~\bibnamefont
  {Vicari}},\ }\href@noop {} {\bibfield  {journal} {\bibinfo  {journal} {Phys.
  Rev. E}\ }\textbf {\bibinfo {volume} {68}},\ \bibinfo {pages} {036136}
  (\bibinfo {year} {2003})}\BibitemShut {NoStop}%
\bibitem [{\citenamefont {Berche}\ \emph {et~al.}(2004)\citenamefont {Berche},
  \citenamefont {Chatelain}, \citenamefont {Berche},\ and\ \citenamefont
  {Janke}}]{Berche2004}%
  \BibitemOpen
  \bibfield  {author} {\bibinfo {author} {\bibfnamefont {P.~E.}\ \bibnamefont
  {Berche}}, \bibinfo {author} {\bibfnamefont {C.}~\bibnamefont {Chatelain}},
  \bibinfo {author} {\bibfnamefont {B.}~\bibnamefont {Berche}}, \ and\ \bibinfo
  {author} {\bibfnamefont {W.}~\bibnamefont {Janke}},\ }\href@noop {}
  {\bibfield  {journal} {\bibinfo  {journal} {Eur. Phys. J. B}\ }\textbf
  {\bibinfo {volume} {38}},\ \bibinfo {pages} {463} (\bibinfo {year}
  {2004})}\BibitemShut {NoStop}%
\bibitem [{\citenamefont {Hasenbusch}\ \emph
  {et~al.}(2007{\natexlab{a}})\citenamefont {Hasenbusch}, \citenamefont
  {Toldin}, \citenamefont {Pelissetto},\ and\ \citenamefont {Vicari}}]{HTPV07}%
  \BibitemOpen
  \bibfield  {author} {\bibinfo {author} {\bibfnamefont {M.}~\bibnamefont
  {Hasenbusch}}, \bibinfo {author} {\bibfnamefont {F.~P.}\ \bibnamefont
  {Toldin}}, \bibinfo {author} {\bibfnamefont {A.}~\bibnamefont {Pelissetto}},
  \ and\ \bibinfo {author} {\bibfnamefont {E.}~\bibnamefont {Vicari}},\
  }\href@noop {} {\bibfield  {journal} {\bibinfo  {journal} {J. Stat. Mech.:
  Theory Exp.}\ }\textbf {\bibinfo {volume} {2007}},\ \bibinfo {pages} {P02016}
  (\bibinfo {year} {2007}{\natexlab{a}})}\BibitemShut {NoStop}%
\bibitem [{\citenamefont {Hasenbusch}\ \emph
  {et~al.}(2007{\natexlab{b}})\citenamefont {Hasenbusch}, \citenamefont
  {Toldin}, \citenamefont {Pelissetto},\ and\ \citenamefont
  {Vicari}}]{PhysRevB.76.094402}%
  \BibitemOpen
  \bibfield  {author} {\bibinfo {author} {\bibfnamefont {M.}~\bibnamefont
  {Hasenbusch}}, \bibinfo {author} {\bibfnamefont {F.~P.}\ \bibnamefont
  {Toldin}}, \bibinfo {author} {\bibfnamefont {A.}~\bibnamefont {Pelissetto}},
  \ and\ \bibinfo {author} {\bibfnamefont {E.}~\bibnamefont {Vicari}},\
  }\href@noop {} {\bibfield  {journal} {\bibinfo  {journal} {Phys. Rev. B}\
  }\textbf {\bibinfo {volume} {76}},\ \bibinfo {pages} {094402} (\bibinfo
  {year} {2007}{\natexlab{b}})}\BibitemShut {NoStop}%
\bibitem [{\citenamefont {Fytas}\ and\ \citenamefont
  {Theodorakis}(2010)}]{PhysRevE.82.062101}%
  \BibitemOpen
  \bibfield  {author} {\bibinfo {author} {\bibfnamefont {N.~G.}\ \bibnamefont
  {Fytas}}\ and\ \bibinfo {author} {\bibfnamefont {P.~E.}\ \bibnamefont
  {Theodorakis}},\ }\href@noop {} {\bibfield  {journal} {\bibinfo  {journal}
  {Phys. Rev. E}\ }\textbf {\bibinfo {volume} {82}},\ \bibinfo {pages} {062101}
  (\bibinfo {year} {2010})}\BibitemShut {NoStop}%
\bibitem [{\citenamefont {Theodorakis}\ and\ \citenamefont
  {Fytas}(2011)}]{Theodorakis2011}%
  \BibitemOpen
  \bibfield  {author} {\bibinfo {author} {\bibfnamefont {P.~E.}\ \bibnamefont
  {Theodorakis}}\ and\ \bibinfo {author} {\bibfnamefont {N.~G.}\ \bibnamefont
  {Fytas}},\ }\href@noop {} {\bibfield  {journal} {\bibinfo  {journal} {Eur.
  Phys. J. B}\ }\textbf {\bibinfo {volume} {81}},\ \bibinfo {pages} {245}
  (\bibinfo {year} {2011})}\BibitemShut {NoStop}%
\bibitem [{\citenamefont {Papakonstantinou}\ and\ \citenamefont
  {Malakis}(2013)}]{PhysRevE.87.012132}%
  \BibitemOpen
  \bibfield  {author} {\bibinfo {author} {\bibfnamefont {T.}~\bibnamefont
  {Papakonstantinou}}\ and\ \bibinfo {author} {\bibfnamefont {A.}~\bibnamefont
  {Malakis}},\ }\href@noop {} {\bibfield  {journal} {\bibinfo  {journal} {Phys.
  Rev. E}\ }\textbf {\bibinfo {volume} {87}},\ \bibinfo {pages} {012132}
  (\bibinfo {year} {2013})}\BibitemShut {NoStop}%
\bibitem [{\citenamefont {Belanger}(2000)}]{B00}%
  \BibitemOpen
  \bibfield  {author} {\bibinfo {author} {\bibfnamefont {D.}~\bibnamefont
  {Belanger}},\ }\href@noop {} {\bibfield  {journal} {\bibinfo  {journal}
  {Braz. J. Phys.}\ }\textbf {\bibinfo {volume} {30}},\ \bibinfo {pages} {682}
  (\bibinfo {year} {2000})}\BibitemShut {NoStop}%
\bibitem [{\citenamefont {Folk}\ \emph {et~al.}(2003)\citenamefont {Folk},
  \citenamefont {Holovatch},\ and\ \citenamefont {Yavorskii}}]{FHY2003}%
  \BibitemOpen
  \bibfield  {author} {\bibinfo {author} {\bibfnamefont {R.}~\bibnamefont
  {Folk}}, \bibinfo {author} {\bibfnamefont {Y.}~\bibnamefont {Holovatch}}, \
  and\ \bibinfo {author} {\bibfnamefont {T.}~\bibnamefont {Yavorskii}},\
  }\href@noop {} {\bibfield  {journal} {\bibinfo  {journal} {Usp. Fiz. Nauk}\
  }\textbf {\bibinfo {volume} {173}},\ \bibinfo {pages} {175} (\bibinfo {year}
  {2003})},\ \bibinfo {note} {[Phys. Uspekhi 46, 169 (2003)]}\BibitemShut
  {NoStop}%
\bibitem [{\citenamefont {Prudnikov}\ \emph {et~al.}(2007)\citenamefont
  {Prudnikov}, \citenamefont {Prudnikov}, \citenamefont {Vakilov},\ and\
  \citenamefont {Krynitsyn}}]{PPVK2007}%
  \BibitemOpen
  \bibfield  {author} {\bibinfo {author} {\bibfnamefont {V.~V.}\ \bibnamefont
  {Prudnikov}}, \bibinfo {author} {\bibfnamefont {P.~V.}\ \bibnamefont
  {Prudnikov}}, \bibinfo {author} {\bibfnamefont {A.~N.}\ \bibnamefont
  {Vakilov}}, \ and\ \bibinfo {author} {\bibfnamefont {A.~S.}\ \bibnamefont
  {Krynitsyn}},\ }\href@noop {} {\bibfield  {journal} {\bibinfo  {journal} {Zh.
  Eksp. Teor. Fiz.}\ }\textbf {\bibinfo {volume} {132}},\ \bibinfo {pages}
  {417} (\bibinfo {year} {2007})},\ \bibinfo {note} {[Phys. JETP 105, 371
  (2007)]}\BibitemShut {NoStop}%
\bibitem [{\citenamefont {Tissier}\ \emph {et~al.}(2002)\citenamefont
  {Tissier}, \citenamefont {Mouhanna}, \citenamefont {Vidal},\ and\
  \citenamefont {Delamotte}}]{Tissier_2002}%
  \BibitemOpen
  \bibfield  {author} {\bibinfo {author} {\bibfnamefont {M.}~\bibnamefont
  {Tissier}}, \bibinfo {author} {\bibfnamefont {D.}~\bibnamefont {Mouhanna}},
  \bibinfo {author} {\bibfnamefont {J.}~\bibnamefont {Vidal}}, \ and\ \bibinfo
  {author} {\bibfnamefont {B.}~\bibnamefont {Delamotte}},\ }\href@noop {}
  {\bibfield  {journal} {\bibinfo  {journal} {Phys. Rev. B}\ }\textbf {\bibinfo
  {volume} {65}},\ \bibinfo {pages} {140402} (\bibinfo {year}
  {2002})}\BibitemShut {NoStop}%
\bibitem [{\citenamefont {Simmons-Duffin}(2017)}]{Simmons-Duffin2017}%
  \BibitemOpen
  \bibfield  {author} {\bibinfo {author} {\bibfnamefont {D.}~\bibnamefont
  {Simmons-Duffin}},\ }\href@noop {} {\bibfield  {journal} {\bibinfo  {journal}
  {J. High Energy Phys.}\ }\textbf {\bibinfo {volume} {2017}},\ \bibinfo
  {pages} {86} (\bibinfo {year} {2017})}\BibitemShut {NoStop}%
\bibitem [{\citenamefont {Antipin}\ and\ \citenamefont
  {Bersini}(2019)}]{Antipin_2019}%
  \BibitemOpen
  \bibfield  {author} {\bibinfo {author} {\bibfnamefont {O.}~\bibnamefont
  {Antipin}}\ and\ \bibinfo {author} {\bibfnamefont {J.}~\bibnamefont
  {Bersini}},\ }\href@noop {} {\bibfield  {journal} {\bibinfo  {journal} {Phys.
  Rev. D}\ }\textbf {\bibinfo {volume} {100}},\ \bibinfo {pages} {065008}
  (\bibinfo {year} {2019})}\BibitemShut {NoStop}%
\bibitem [{\citenamefont {Stergiou}(2019)}]{Stergiou_2019}%
  \BibitemOpen
  \bibfield  {author} {\bibinfo {author} {\bibfnamefont {A.}~\bibnamefont
  {Stergiou}},\ }\href@noop {} {\bibfield  {journal} {\bibinfo  {journal}
  {SciPost Phys.}\ }\textbf {\bibinfo {volume} {7}},\ \bibinfo {pages} {10}
  (\bibinfo {year} {2019})}\BibitemShut {NoStop}%
\bibitem [{\citenamefont {Kousvos}\ and\ \citenamefont
  {Stergiou}(2019)}]{Kousvos_2019}%
  \BibitemOpen
  \bibfield  {author} {\bibinfo {author} {\bibfnamefont {S.~R.}\ \bibnamefont
  {Kousvos}}\ and\ \bibinfo {author} {\bibfnamefont {A.}~\bibnamefont
  {Stergiou}},\ }\href@noop {} {\bibfield  {journal} {\bibinfo  {journal}
  {SciPost Phys.}\ }\textbf {\bibinfo {volume} {6}},\ \bibinfo {pages} {35}
  (\bibinfo {year} {2019})}\BibitemShut {NoStop}%
\bibitem [{\citenamefont {Nandi}\ and\ \citenamefont
  {T\"auber}(2020)}]{Nandi_2020}%
  \BibitemOpen
  \bibfield  {author} {\bibinfo {author} {\bibfnamefont {R.}~\bibnamefont
  {Nandi}}\ and\ \bibinfo {author} {\bibfnamefont {U.~C.}\ \bibnamefont
  {T\"auber}},\ }\href@noop {} {\bibfield  {journal} {\bibinfo  {journal}
  {Phys. Rev. E}\ }\textbf {\bibinfo {volume} {102}},\ \bibinfo {pages}
  {052114} (\bibinfo {year} {2020})}\BibitemShut {NoStop}%
\bibitem [{\citenamefont {Shapoval}\ \emph {et~al.}(2020)\citenamefont
  {Shapoval}, \citenamefont {Dudka}, \citenamefont {Fedorenko},\ and\
  \citenamefont {Holovatch}}]{Shapoval_2020}%
  \BibitemOpen
  \bibfield  {author} {\bibinfo {author} {\bibfnamefont {D.}~\bibnamefont
  {Shapoval}}, \bibinfo {author} {\bibfnamefont {M.}~\bibnamefont {Dudka}},
  \bibinfo {author} {\bibfnamefont {A.~A.}\ \bibnamefont {Fedorenko}}, \ and\
  \bibinfo {author} {\bibfnamefont {Y.}~\bibnamefont {Holovatch}},\ }\href@noop
  {} {\bibfield  {journal} {\bibinfo  {journal} {Phys. Rev. B}\ }\textbf
  {\bibinfo {volume} {101}},\ \bibinfo {pages} {064402} (\bibinfo {year}
  {2020})}\BibitemShut {NoStop}%
\bibitem [{\citenamefont {Vigneshwar}\ \emph {et~al.}(2019)\citenamefont
  {Vigneshwar}, \citenamefont {Mandal}, \citenamefont {Damle}, \citenamefont
  {Dhar},\ and\ \citenamefont {Rajesh}}]{Vigneshwar_2019}%
  \BibitemOpen
  \bibfield  {author} {\bibinfo {author} {\bibfnamefont {N.}~\bibnamefont
  {Vigneshwar}}, \bibinfo {author} {\bibfnamefont {D.}~\bibnamefont {Mandal}},
  \bibinfo {author} {\bibfnamefont {K.}~\bibnamefont {Damle}}, \bibinfo
  {author} {\bibfnamefont {D.}~\bibnamefont {Dhar}}, \ and\ \bibinfo {author}
  {\bibfnamefont {R.}~\bibnamefont {Rajesh}},\ }\href@noop {} {\bibfield
  {journal} {\bibinfo  {journal} {Phys. Rev. E}\ }\textbf {\bibinfo {volume}
  {99}},\ \bibinfo {pages} {052129} (\bibinfo {year} {2019})}\BibitemShut
  {NoStop}%
\bibitem [{\citenamefont {Adzhemyan}\ \emph {et~al.}(2019)\citenamefont
  {Adzhemyan}, \citenamefont {Ivanova}, \citenamefont {Kompaniets},
  \citenamefont {Kudlis},\ and\ \citenamefont {Sokolov}}]{ADZHEMYAN2019332}%
  \BibitemOpen
  \bibfield  {author} {\bibinfo {author} {\bibfnamefont {L.~T.}\ \bibnamefont
  {Adzhemyan}}, \bibinfo {author} {\bibfnamefont {E.~V.}\ \bibnamefont
  {Ivanova}}, \bibinfo {author} {\bibfnamefont {M.~V.}\ \bibnamefont
  {Kompaniets}}, \bibinfo {author} {\bibfnamefont {A.}~\bibnamefont {Kudlis}},
  \ and\ \bibinfo {author} {\bibfnamefont {A.~I.}\ \bibnamefont {Sokolov}},\
  }\href@noop {} {\bibfield  {journal} {\bibinfo  {journal} {Nucl. Phys. B}\
  }\textbf {\bibinfo {volume} {940}},\ \bibinfo {pages} {332} (\bibinfo {year}
  {2019})}\BibitemShut {NoStop}%
\bibitem [{\citenamefont {Kompaniets}\ and\ \citenamefont
  {Panzer}(2017)}]{KP17}%
  \BibitemOpen
  \bibfield  {author} {\bibinfo {author} {\bibfnamefont {M.~V.}\ \bibnamefont
  {Kompaniets}}\ and\ \bibinfo {author} {\bibfnamefont {E.}~\bibnamefont
  {Panzer}},\ }\href@noop {} {\bibfield  {journal} {\bibinfo  {journal} {Phys.
  Rev. D}\ }\textbf {\bibinfo {volume} {96}},\ \bibinfo {pages} {036016}
  (\bibinfo {year} {2017})}\BibitemShut {NoStop}%
\bibitem [{\citenamefont {Le~Guillou}\ and\ \citenamefont
  {Zinn-Justin}(1980)}]{PhysRevB.21.3976}%
  \BibitemOpen
  \bibfield  {author} {\bibinfo {author} {\bibfnamefont {J.~C.}\ \bibnamefont
  {Le~Guillou}}\ and\ \bibinfo {author} {\bibfnamefont {J.}~\bibnamefont
  {Zinn-Justin}},\ }\href@noop {} {\bibfield  {journal} {\bibinfo  {journal}
  {Phys. Rev. B}\ }\textbf {\bibinfo {volume} {21}},\ \bibinfo {pages} {3976}
  (\bibinfo {year} {1980})}\BibitemShut {NoStop}%
\bibitem [{\citenamefont {Guida}\ and\ \citenamefont
  {Zinn-Justin}(1997)}]{GZ97}%
  \BibitemOpen
  \bibfield  {author} {\bibinfo {author} {\bibfnamefont {R.}~\bibnamefont
  {Guida}}\ and\ \bibinfo {author} {\bibfnamefont {J.}~\bibnamefont
  {Zinn-Justin}},\ }\href@noop {} {\bibfield  {journal} {\bibinfo  {journal}
  {Nucl. Phys. B}\ }\textbf {\bibinfo {volume} {489}},\ \bibinfo {pages} {626}
  (\bibinfo {year} {1997})}\BibitemShut {NoStop}%
\bibitem [{\citenamefont {Guida}\ and\ \citenamefont
  {Zinn-Justin}(1998)}]{GZJ98}%
  \BibitemOpen
  \bibfield  {author} {\bibinfo {author} {\bibfnamefont {R.}~\bibnamefont
  {Guida}}\ and\ \bibinfo {author} {\bibfnamefont {J.}~\bibnamefont
  {Zinn-Justin}},\ }\href@noop {} {\bibfield  {journal} {\bibinfo  {journal}
  {J. Phys. A}\ }\textbf {\bibinfo {volume} {31}},\ \bibinfo {pages} {8103}
  (\bibinfo {year} {1998})}\BibitemShut {NoStop}%
\bibitem [{\citenamefont {Zinn-Justin}(2001)}]{ZJ01}%
  \BibitemOpen
  \bibfield  {author} {\bibinfo {author} {\bibfnamefont {J.}~\bibnamefont
  {Zinn-Justin}},\ }\href@noop {} {\bibfield  {journal} {\bibinfo  {journal}
  {Phys. Reports}\ }\textbf {\bibinfo {volume} {344}},\ \bibinfo {pages} {159}
  (\bibinfo {year} {2001})}\BibitemShut {NoStop}%
\bibitem [{\citenamefont {Zinn-Justin}(2002)}]{ZJ}%
  \BibitemOpen
  \bibfield  {author} {\bibinfo {author} {\bibfnamefont {J.}~\bibnamefont
  {Zinn-Justin}},\ }\href@noop {} {\emph {\bibinfo {title} {Quantum Field
  Theory and Critical Phenomena}}}\ (\bibinfo  {publisher} {Oxford University
  Press},\ \bibinfo {year} {2002})\BibitemShut {NoStop}%
\bibitem [{\citenamefont {Pelissetto}\ and\ \citenamefont
  {Vicari}(2002)}]{PV02}%
  \BibitemOpen
  \bibfield  {author} {\bibinfo {author} {\bibfnamefont {A.}~\bibnamefont
  {Pelissetto}}\ and\ \bibinfo {author} {\bibfnamefont {E.}~\bibnamefont
  {Vicari}},\ }\href@noop {} {\bibfield  {journal} {\bibinfo  {journal} {Phys.
  Reports}\ }\textbf {\bibinfo {volume} {368}},\ \bibinfo {pages} {549}
  (\bibinfo {year} {2002})}\BibitemShut {NoStop}%
\bibitem [{\citenamefont {Ma}(1976)}]{MaBook}%
  \BibitemOpen
  \bibfield  {author} {\bibinfo {author} {\bibfnamefont {S.~K.}\ \bibnamefont
  {Ma}},\ }\href@noop {} {\emph {\bibinfo {title} {Modern theory of critical
  phenomena}}}\ (\bibinfo  {publisher} {Benjamin-Cummings, Reading, MA},\
  \bibinfo {year} {1976})\BibitemShut {NoStop}%
\end{thebibliography}%
\end{document}